\documentclass[manuscript]{acmart}
\AtBeginDocument{%
  \providecommand\BibTeX{{%
    \normalfont B\kern-0.5em{\scshape i\kern-0.25em b}\kern-0.8em\TeX}}}

\copyrightyear{2025}
\acmYear{2025}
\acmConference[CHI '25]{CHI Conference on Human Factors in Computing Systems}{April 26-May 1, 2025}{Yokohama, Japan}
\acmBooktitle{CHI Conference on Human Factors in Computing Systems (CHI '25), April 26-May 1, 2025, Yokohama, Japan}\acmDOI{10.1145/3706598.3713338}
\acmISBN{979-8-4007-1394-1/25/04}

\newcommand{\msd}[2]{mean = #1, $SD$ = #2}
\newcommand{\ttest}[3]{$t(#1)$ = $#2$, $p$ < #3}
\newcommand{\ttestno}[3]{$t(#1)$ = $#2$, $p$ = #3}

\newcommand{\odds}[3]{OR = #1, $z$ = #2, p < #3}
\newcommand{\diff}[1]{diff = $#1$}
\newcommand{\repo}[1]{\url{https://tinyurl.com/GenAICopyrightOpinions}}

\usepackage{graphicx,subfigure,enumitem}
\usepackage{bm}

\begin{document}

\title[Public Opinions About Copyright for AI-Generated Art]{Public Opinions About Copyright for AI-Generated Art: \\The Role of Egocentricity, Competition, and Experience}

\author{Gabriel Lima}
\affiliation{%
  \institution{Max Planck Institute for Security and Privacy}
  \country{Germany}}
\email{gabriel.lima@mpi-sp.org}

\author{Nina Grgi\'{c}-Hla\v{c}a}
\affiliation{%
  \institution{Max Planck Institute for Software Systems \& Max Planck Institute for Research on Collective Goods}
  \country{Germany}
}
\email{nghlaca@mpi-sws.org}

\author{Elissa M. Redmiles}
\affiliation{%
  \institution{Georgetown University}
  \country{United States}
}
\email{elissa.redmiles@georgetown.edu}


\begin{abstract}
    
Breakthroughs in generative AI (GenAI) have fueled debates concerning the artistic and legal status of AI-generated creations. We investigate laypeople's perceptions ($N$$=$$432$) of AI-generated art through the lens of copyright law. We study lay judgments of GenAI images concerning several copyright-related factors and capture people's opinions of who should be the authors and rights-holders of AI-generated images. To do so, we held an incentivized AI art competition in which some participants used a GenAI model to create art while others evaluated these images. We find that participants believe creativity and effort, but not skills, are needed to create AI-generated art. Participants were most likely to attribute authorship and copyright to the AI model's users and to the artists whose creations were used for training. We find evidence of egocentric effects: participants favored their own art with respect to quality, creativity, and effort---particularly when these assessments determined real monetary awards.

\end{abstract}


\keywords{Generative AI, Large Language Models, GenAI, LLM, Copyright, Egocentric Effects, Competition, Exhibition, Art, AI-Generated Art, Intellectual Property, IP}


\maketitle

\section{Introduction}

\label{sec:intro}

Recent breakthroughs in generative artificial intelligence (GenAI) have pushed the boundaries of what machines can generate across various domains. Text-based models~\cite{achiam2023gpt} have become pervasive online, enabling users to generate long-form texts from short prompts and revolutionizing how one searches for information online. Similarly, AI image models~\cite{betker2023improving} have empowered their users to generate highly realistic and detailed images from short textual descriptions. Multimodal GenAI models~\cite{team2023gemini} show even more promise, generating text, images, audio, and other types of data. These systems learn from large datasets of human creations how to generate text, images, and other content that may be indistinguishable from their human-created counterparts~\cite{kobis2021artificial,longoni2022news,jakesch2023human}. {GenAI models have now become part of several real-world products~\cite{mediumnail}, such as search engines~\cite{googleGenAI}, online shopping websites~\cite{amazonGenAI}, and messaging apps~\cite{whatsappAI}.}

{Although GenAI models have the potential to revolutionize how humans interact with technology and express their creativity, they also pose novel challenges to society.} For instance, there exist debates concerning whether GenAI outputs could devalue human creativity~\cite{merkley}, fuel artistic appropriation~\cite{goetze2024ai}, and concentrate power in the hands of a few corporate actors~\cite{iaia2022or,chan2023reclaiming}. Such debates are both philosophical---concerning aesthetics and societal perceptions of what art is and is not---and legal. Copyright law, the body of law that regulates works of authorship (e.g., paintings and novels) and determines who should have exclusive rights over these creations, has received considerable attention in the context of GenAI. Extensive literature has examined how copyright law should address AI-generated works (e.g., ~\cite{guadamuz2017androids,gervais2019machine,smits2022generative,eshraghian2020human,lee2023talkin}, and several lawsuits are currently underway~\cite{nytlawsuit,copyrightverge} to determine whether training on copyrighted material violates the law and if AI-generated content warrants the same legal protection as human-created works. 

Thus far, considerations of AI-generated outputs from a copyright lens have been primarily normative (e.g.,~\cite{guadamuz2017androids,gervais2019machine}), with little focus on capturing the opinion of GenAI users. Scholars debate whether AI-generated outputs are eligible for copyright protection~\cite{smits2022generative} and, if so, who should hold the rights associated with this protection~\cite{eshraghian2020human}. Here, we investigate \emph{laypeople's perceptions} of GenAI art through a copyright lens. 

\begin{figure*}[ht!]
    \centering
    \includegraphics[width=0.7\textwidth]{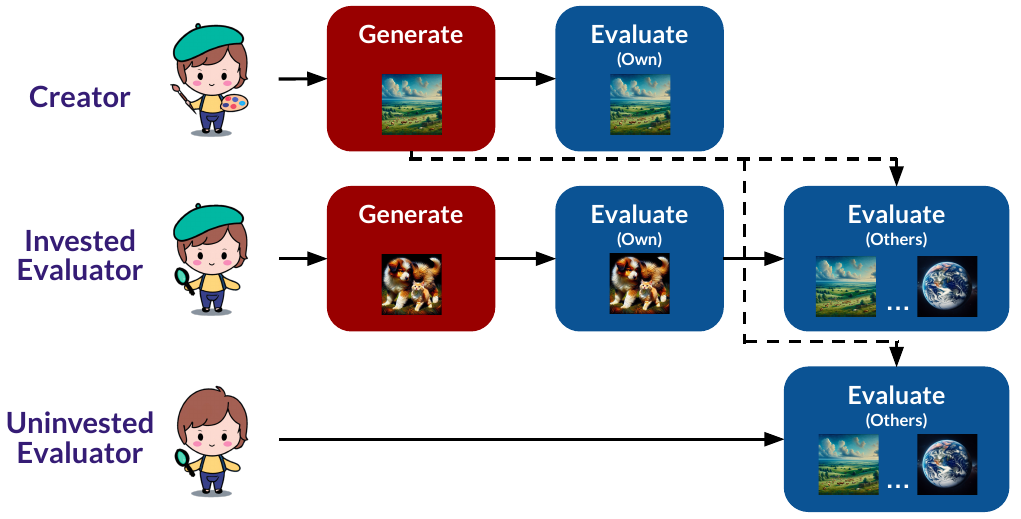}
    \caption{High-level overview of our experiment.}
    \Description{
    The figure is a flowchart depicting the three experimental conditions: creator, invested evaluator, and uninvested evaluator. Creators generate an image for consideration and then evaluate their own creation. Invested evaluators generate an image, evaluate their own creation, and then evaluate images generated by creators. Uninvested evaluators only evaluate images generated by creators.}
    \label{fig:methods}
\end{figure*}

There are several reasons why studying lay opinions concerning the intersection of copyright and GenAI is relevant. First, capturing laypeople's perceptions of the law more generally is important to ensure that it is democratically legitimate~\cite{tobia2022experimental}. Democratic theories of law argue that the law should reflect laypeople's intuitions~\cite{tobia2022experimental} to motivate citizens to comply with it~\cite{tyler2014popular}. This paper's approach can thus help ensure that future legal decisions and policymaking are aligned with public expectations. Even if the law is at odds with lay intuitions, capturing laypeople's opinions can help mitigate any potential backlash that may emerge from these differences, proposing ways to bridge these gaps~\cite{awad2020crowdsourcing}.

Second, aligning laypeople's legal intuitions with the law is particularly important in the context of copyright. Copyright law has several objectives~\cite{lee2023talkin}, such as promoting fairness~\cite{damich1988right} and safeguarding creators' moral rights~\cite{damich1988right}. Copyright also aims to incentivize creativity through financial incentives. By granting exclusive rights over creative works to authors, copyright law attempts to incentivize them to continue exercising their creativity for further financial benefit~\cite{mandel2014public,mandel2016ip}. Because copyright law depends on this behavioral response to achieve its objective (i.e., to promote creativity), it is all-important that potential creators, namely laypeople, understand how the law works~\cite{mandel2014public}. 

Third, empirical studies capturing the opinion of \emph{laypeople} about copyright are distinctively important in the context of GenAI. GenAI aims to democratize the ability to create works that could be eligible for copyright protection---something that used to be restricted to skilled artists---enabling non-artists to participate in the creative society.

{Research on how novel digital systems impact people's perceptions of ownership and shape individuals' processes for meaning-making around the digital artifacts produced by those same systems is core to human-computer interaction (HCI)~\cite{xu2020you,gulotta2013digital}. We take the law as a perspective lens through which we explore people's perceptions of one of the digital artifacts of GenAI systems: AI-generated art. Because regulations constraint what users and HCI designers can do, capturing people's perceptions of technology through the lens of the law sheds light on how expectations might clash with policy~\cite{awad2020crowdsourcing} and provides implications for designers aiming to develop systems that function within legal constraints while satisfying user needs~\cite{fiesler2020lawful}. In view of the recent developments in GenAI and how this technology has become embedded in both HCI research and real-world products, our research aims to inform the HCI community towards the design of GenAI systems that are both lawful and fulfilling to users~\cite{lazar2019introduction}.}

{We acknowledge that debates on how novel technologies can change the meaning of existing cultural artifacts and their regulation are not a new development. The rise of photography also challenged what was eligible for copyright protection~\cite{hughes2011photographer}. Similarly, the law was initially unprepared to deal with the emergence of digital art~\cite{geller2000copyright} and video-sharing platforms that allowed users to upload copyrighted content without much restriction, leading to solutions that ensure that the creator rights are protected (e.g., Google's content ID~\cite{contentID}). Now,  GenAI poses novel policy questions by contesting what warrants (or not) copyright protection---the topic of the research we present below.}

\subsection{{Empirical Study of AI-Generated Art Through the Lens of Copyright}} 
This paper examines laypeople's perceptions ($N$ = 432) of AI-generated images in relation to their potential copyright protection. We first capture how laypeople judge GenAI images with respect to factors that help determine whether human creations are eligible for copyright in different jurisdictions~\cite{eshraghian2020human} (see Section~\ref{sec:factors} for more detailed legal background). Then, we investigate whether laypeople believe AI-generated images warrant copyright and, if so, who should own them (see Section~\ref{sec:ownership} for more detailed legal background). More specifically, we address the following research questions:

\begin{itemize}[label={},leftmargin=*, itemsep=5pt]
    \item \textbf{RQ1:} How do laypeople evaluate AI-generated images concerning the creativity, effort, and skills involved in the creation process? 
    \item \textbf{RQ2:} Who do laypeople believe are the authors of AI-generated images?
    \item \textbf{RQ3:} Who do laypeople believe should hold the rights to 1) display and 2) make copies of AI-generated images?
\end{itemize}

To investigate laypeople’s perceptions of AI-generated art, we conducted an experimental study in the form of a juried AI art exhibition. To guide our experimental design, we leveraged prior research that found that people judge their own creations (or the creations they own) differently than the creations of others---the literature often terms this ``egocentric bias''~\cite{norton2012ikea,buccafusco2011creativity,buccafusco2010valuing,spellman2008artists}. We hypothesize that similar biases may emerge in people's perceptions of copyright since they relate to questions surrounding ownership of creative works. 

Participants engaged in the exhibition either as creators (by using a GenAI model to create art), invested evaluators (by generating art and evaluating other people's submissions), or uninvested evaluators (by only evaluating others' images). Figure~\ref{fig:methods} presents a high-level overview of our experimental design. Our between-subjects design allowed us to study how perceptions about copyright vary between those who have something to gain from copyright protection and the exhibition---creators of AI-generated art---and uninvested third parties. For instance, we hypothesized that creators of AI-generated art would egocentrically overestimate the quality of their own creations and exhibit greater support for the exclusive ownership structure that copyright protection could afford them.

Our study was designed to maximize ecological validity. The decision to hold an AI art exhibition was inspired by real-world examples of AI-generated images winning art exhibitions~\cite{nytexhibition} and competitions focused solely on AI art~\cite{smithexhibtion}. Our AI art exhibition rewarded the top-10 best submissions, mimicking some of the financial incentives involved in copyright decisions. It also simulates other non-monetary incentives, such as exposure and recognition; selected images were displayed on a website, in which participants had the choice to attach their names to their creations.\footnote{\url{https://thegcamilo.github.io/AI-art-exhibition/}}

\subsection{Findings \& Implications}
Our results suggest that people believe creativity and effort---but not necessarily skill---are necessary to create art using AI (RQ1). Participants also indicated that users and those whose creations were used to train the GenAI model should be considered authors of AI-generated images (RQ2) and enjoy the rights to display and make copies of them (RQ3). In contrast, people were less likely to attribute authorship (RQ2) and the rights associated with copyright protection (RQ3) to the AI model itself and the company that developed it.

Our research has implications for the development of GenAI models and their future regulation under copyright law. Our findings call for the consideration of a more distributed ownership structure of copyright, under which training data contributors are also recognized as authors and rights-holders. People's attribution of authorship and rights to data contributors rather than the company that developed the AI model raises questions concerning current business models that concentrate profits in corporate entities at the expense of human artists~\cite{stealingnewyorker,merkley}. {Our findings interrogate current practices of GenAI model designers, who may---knowingly or accidentally---promote the interests of powerful corporate actors to the detriment of artists' welfare. We discuss how existing legal doctrines (e.g., neighbouring rights, licensing models), computer science research, and self-questioning from those who design GenAI products could help ensure that training data contributors are compensated.}

Focusing on how people's perceptions vary egocentrically, we found that participants evaluate the process of using GenAI to generate art egocentrically with respect to some factors---e.g., creativity and effort---but not others---such as skills. Our results indicate that egocentric biases become particularly relevant when monetary incentives come into play. Although participants did not prioritize themselves when asked to indicate who should hold \emph{hypothetical} rights over their creations, they overestimated the quality of their images when that determined \emph{real} monetary rewards. Namely, when deciding who should win the art exhibition's monetary award, participants judged their own creations much more favorably than art generated by others, supporting our egocentric hypotheses. Surprisingly, we found the opposite trend in attributions of authorship to the user of the GenAI model. Creators were less likely to rate themselves as authors than evaluators, potentially suggesting that GenAI users are aware that authorship can bring both benefits and liability for their creations. Our evidence of egocentric biases suggests that some conflicts of interest may arise in discussions surrounding the legal status of AI-generated art under copyright law. {We also discuss how decisions concerning the terms of use of real-world GenAI systems may shape people's expectations of AI-generated images in light of their potential financial returns.}

\section{Background}
\label{sec:background}
Generative AI (GenAI) has the potential to revolutionalize how humans exercise their creativity. However, it does not come without problems. Several reports indicate that GenAI has the potential to amplify harmful stereotypes~\cite{stereotypepost,stereotypebloomberg}. Researchers have also warned how generative models could fuel online mis- and disinformation by generating false, yet credible-looking, news and online profiles~\cite{augenstein2023factuality}. These models' tendency to fabricate information while sounding confident and knowledgeable can also distort human beliefs~\cite{kidd2023ai}. There exists evidence that GenAI can produce misinformation that is more compelling to readers~\cite{spitale2023ai}, as well as manipulate people's beliefs in conspiracy theories~\cite{costello2024durably}.

One particular domain that has been directly impacted by the emergence of GenAI is intellectual property (IP) law. IP law refers to the rules that regulate the rights associated with human creations, such as inventions and literary and artistic works, determining who should control and benefit from them. How to deal with AI-generated outputs regarding IP rights remains an open question. For instance, should machine-generated artistic works be protected similarly to their human-created counterparts? Who should enjoy the rights that would normally be associated with these creations?

In this paper, we focus on copyright law, which is the branch of IP law that covers works of authorship, including artistic, musical, and literary works, such as novels, movies, songs, and many other human creations.\footnote{Discussions surrounding GenAI and IP law are not restricted to copyright. For instance, some have argued that AI should be treated as an inventor under patent law~\cite{abbott2020reasonable}, which deals with the rights associated with human inventions. This argument has been successful in Australia~\cite{patentwipo} and, at the same time, has found challenges in the United States~\cite{patentverge}.} Although what qualifies for copyright protection and what rights follow this determination vary by country, copyright law mainly grants some exclusive rights (e.g., to reproduction and distribution) to the copyright holder for a predetermined period of time. In the case of human creations, copyright owners are often the creators themselves, with exceptions in case the work has been created within the scope of employment.

The challenges posed by GenAI to copyright law can be grouped in three overarching questions~\cite{franceschelli2022copyright}: 1) does training AI models on copyrighted data infringe on the copyright of the training data?; 2) are AI-generated outputs eligible for copyright protection; and if so 3) who owns the copyright? {In this paper, we focus on the two latter questions, both of which we discuss in \S\ref{sec:factors}-\ref{sec:ownership} below.\footnote{{We also present the debate on how training GenAI models might violate the copyright of the training data (i.e., question \#1 above), as well as some potential alternative regulatory frameworks for GenAI in the Appendix for conciseness.}} Furthermore, we present prior work on lay perceptions of copyright (\S\ref{sec:lay_copyright}) and GenAI (\S\ref{sec:lay_genAI}) and motivate our focus on potential biases that may emerge in the context of copyright and GenAI (\S\ref{sec:biases}).}

\subsection{Can AI-Generated Outputs Be Protected by Copyright?}
\label{sec:factors}

An important question raised by GenAI concerning copyright law refers to whether AI-generated outputs should receive the same protection that is assigned to works that are created solely by humans. If works generated with the assistance of an AI model are eligible for copyright protection, someone would have exclusive rights over it; on the other hand, if they are not eligible, these outputs would be part of the public domain, meaning that anyone would be able to use these works without permission. 

Scholars disagree on this particular question. Some argue extending copyright law to machine-created works would reduce the value of human creativity~\cite{merkley}, flood the market with creations of questionable quality~\cite{gervais2019machine}, and concentrate power in the hands of a few~\cite{iaia2022or,chan2023reclaiming}. In contrast, proponents of extending copyright to AI-generated works defend that protecting these outputs could promote innovation by incentivizing research and development of AI~\cite{hristov2016artificial,kop2019ai}, as well as enable users to create works that would not be possible without GenAI.

Current legal decisions addressing whether AI-generated outputs should be granted copyright have varied widely across different countries. While a judge in the United States (US) has ruled that AI-generated images are not eligible for copyright protection~\cite{copyrightverge}, the Beijing Internet Court has taken a different approach by granting copyright to AI-generated art~\cite{chinacopyright}. Even within China, different jurisdictions have made conflicting copyright decisions regarding GenAI outputs~\cite{wan2021copyright}. These rulings demonstrate how the requirements for copyright protection, and hence the answer to whether AI-generated outputs are eligible, may vary depending on the jurisdiction~\cite{eshraghian2020human,selvadurai2020reconsidering}. While the US Copyright Office requires originality for copyright protection, meaning that a work must be independently created and exhibit a modicum of creativity~\cite{uscopyrightlaw}, the EU posits a work is eligible if it is the result of the ``author's own intellectual creation''~\cite{smits2022generative,hugenholtz2021copyright}. In contrast, other countries (e.g., Australia~\cite{1964ladbroke} and Canada~\cite{2004cch}) also consider whether there was a non-trivial exercise of skill and effort~\cite{eshraghian2020human,guadamuz2017androids}. The United Kingdom (UK) is one of the few countries with clear rules for ``computer-generated works,'' granting exclusive rights to ``the person by whom the arrangements necessary for the creation of the work are undertaken''~\cite{ukcopyright}. \textbf{Motivated by these conflicting viewpoints, this paper explores how laypeople judge AI-generated images with respect to the creativity, effort, and skills involved in the creation process (RQ1)}.

\subsection{Who Owns the Potential Copyright of AI-Generated Outputs?}
\label{sec:ownership}

If an AI-generated work is granted copyright, some legal entity would have exclusive rights over it; but who would that be? A GenAI output is the result of a collaboration between several actors, including the model's developers, its users, those who potentially own the training data, and the AI model itself, making it difficult to determine who should own it.\footnote{This question is an instance of the ``problem of many hands''~\cite{van2015problem}, which posits that it is difficult to determine who is ultimately responsible for collective actions. Scholars have also explored how AI may complicate this search for a responsible actor, particularly when it causes harm~\cite{coeckelbergh2020artificial,matthias2004responsibility}.}

A common proposition is that the user of a GenAI model, i.e., the person who gave it instructions, should be granted exclusive rights over its outputs~\cite{eshraghian2020human}. However, it is not clear whether this user would satisfy the conditions that determine authorship and, hence, copyright ownership. The US Copyright Office has stated that merely prompting an AI model does not qualify the user for authorship; instead, it proposes a case-by-case analysis that would determine whether the work contains ``sufficient human authorship'' (e.g., whether the user ``select[ed] or arrange[d] AI-generated material in a sufficiently creative way''~\cite{us2023copyright}). Another possibility would be granting copyright to the ``the person by whom the arrangements necessary for the creation of the work are undertaken,'' as proposed by the UK Intellectual Property Office~\cite{ukcopyright}, which could interpreted as the developer of the AI model~\cite{ihalainen2018computer}. Other proposals include considering the AI model itself as an author (and thus copyright holder)~\cite{smits2022generative,lima2020collecting} or granting a form of joint authorship (i.e., collective ownership) to the many entities involved~\cite{khosrowi2023diffusing}. \textbf{This research investigates whom laypeople consider to be authors of AI-generated images (RQ2), as well as who they believe should have the rights to display and make copies of these images (RQ3).}

\subsection{Lay Perceptions of Copyright}
\label{sec:lay_copyright}

Copyright protection exists to achieve several objectives~\cite{lee2023talkin}. For instance, it promotes fairness by granting authors the right to exclusively control the fruits of their own labor~\cite{bair2017rational}. It also safeguards the moral rights of creators, protecting the emotional bond between authors and their works~\cite{damich1988right}. Most relevant to this research is copyright law's aim to incentivize creativity. Copyright law attempts to promote creativity by granting exclusive rights over creative works to authors. These exclusive rights determine that only authors can profit from their creations, incentivizing them to continue exercising their creativity for further financial benefit~\cite{mandel2014public,mandel2016ip}. Unless potential authors and rights-holders (i.e., laypeople) understand and agree with what is eligible for copyright protection and what rights are associated with it, copyright law may fail to incentivize the production of creative outputs. Hence, examining public perceptions and expectations of IP law is essential to ensure that current regulations effectively meet their goals and to identify potential changes if necessary.

Prior work looking at perceptions of IP law and authorship demonstrates that lay perceptions may be in conflict with what the law proposes. Laypeople perceive IP law's main objective as preventing plagiarism, although its main aim is more utilitarian by promoting creativity through exclusive rights~\cite{mandel2016ip}. In the internet, content creators also have mistaken beliefs about the copyright terms of the platforms they use~\cite{fiesler2016reality}. \textbf{This paper investigates laypeople's perceptions of visual art generated with the assistance of a GenAI model through the lens of copyright law.}

\subsection{Lay Perceptions of GenAI Outputs}
\label{sec:lay_genAI}

Recent work has explored how people use and perceive GenAI models and their outputs. While some report that people have an inherent bias against AI-generated text and images~\cite{gu2022made}, particularly among those with stronger anthropocentric beliefs in order to ``protect'' human creativity~\cite{millet2023defending}, other studies have found that laypeople prefer human-created works because AI outputs are perceived to be of lower-quality~\cite{bower2021perceptions,kobis2021artificial}. At the same time, many studies demonstrate that people are bad at distinguishing between human- and AI-generated images and text~\cite{kobis2021artificial,longoni2022news,mink2022deepphish}, potentially because of flawed heuristics used to determine whether something is machine-generated~\cite{jakesch2023human,mink2024s}. 

Research has also explored perceptions of GenAI outputs in relation to their ownership and authorship, albeit not in direct relation to their potential copyright protection. Human creators assisted by GenAI are attributed more credit than creators working alongside another human~\cite{jago2023made}. An experiment found that perceptions of authorship of the AI model and related actors can be manipulated by how anthropomorphized the machine is~\cite{epstein2020gets}. Similarly, one's perceived ownership over a piece of text they generated with AI assistance was found to vary depending on the AI's writing style~\cite{kadoma2023role}. 

A few studies have also investigated people's perceptions about issues related to copyright in the context of AI. A study that examined public perceptions about the possibility of granting AI various rights suggests that people may not be contrary to the idea of granting copyright rights to the AI model itself~\cite{lima2020collecting}. A poll on people's thoughts, feelings, and fears about AI found that laypeople believe artists whose creations are being used to train GenAI should be compensated~\cite{vergesurvey}.
\textbf{We build upon this prior work to comprehensively investigate people's opinions regarding GenAI through the lens of copyright law, providing implications to the development and regulation of GenAI models.}

We note that the question of copyright and authorship can also be framed as a question of responsibility. For instance, who is responsible for an AI-generated work and, thus, should have the corresponding rights? Although extensive literature has explored how laypeople attribute responsibility for \emph{harms} caused by AI~\cite{lima2021conflict,lima2021human,lima2023blaming,lee2021computer,malle2015sacrifice,kim2006should}, it is still an open question whether the results would replicate in the case of positive responsibility, such as attributing credit for AI-generated works~\cite{porsdam2023generative,danaher2021automation}, which we consider in this work.

\subsection{Effects of
Egocentricity, Competition, and Experience on Perceptions of Copyright}
\label{sec:biases}

{One's attitudes and beliefs towards a wide range of issues can be biased egocentrically. For instance, people judge fairness violations against themselves more harshly than similar transgressions against others~\cite{greenberg1983overcoming, thompson1992egocentric}. Furthermore, individuals believe they know more about others than others know about them \cite{pronin2001you}, expect mass media to have a larger influence on others than on themselves~\cite{davison1983third}, and even assume that others are more susceptible to egocentric bias than they are~\cite{kruger1999naive}.}

{Particularly relevant for our study, people exhibit egocentric biases in their perceptions of creative works. Authors of creative works overestimate the value of their products~\cite{buccafusco2010valuing}. Similarly, people value their self-made products as much as experts' creations and expect others to share this view~\cite{norton2012ikea}. Hence, people may value AI-generated art they create more than they value others' art and more than others value their art. Moreover, individuals overestimate the creativity of their own creations~\cite{buccafusco2011creativity} and their own contributions to group projects~\cite{ross1979egocentric}, suggesting that similar egocentric biases may emerge in judgments of creativity and authorship, which help determine whether a creation is eligible for copyright protection~\cite{hugenholtz2021copyright,1991feist}. Combined, this body of research motivates us to study \textbf{egocentric effects} on people's perceptions of copyright of AI-generated art.} 

{Another bias that may emerge in the context of our study is what we call the \textbf{competition effect}. In our study, the highest-scored submissions received monetary rewards, meaning that participants' outcomes depended on other people's assessments. Some of those competing for the reward had the opportunity to inflate their own relative rating not only by giving high ratings to their own submission (in line with egocentric effects), but also by giving low scores to other people's submissions. That is, they could benefit from sabotaging their competition. The problem of sabotage in competitions has been discussed extensively in economics: theory predicts its occurrence, and experiments show that sabotage is indeed empirically relevant~\cite{chowdhury2015sabotage}. Hence, creators of AI-generated art may sabotage others for their own financial interests. The competition effect is relevant in the context of copyright law because the financial interests it aims to protect might be susceptible to sabotage when competitors can determine whether something is eligible for copyright protection, as explored in our study. Therefore, we hypothesize that the incentive structure associated with our study may lead those with something at stake to judge others' submissions more harshly than those who do not have anything to gain from doing so.}

{Finally, the experience of interacting with the GenAI model to create AI-generated art could impact one's perceptions of outputs. Research has identified several psychological phenomena, such as the mere-exposure effect~\cite{zajonc1968attitudinal, montoya2017re}, practice and learning effects~\cite{donovan1999meta}, and the curse of knowledge~\cite{hinds1999curse}, which lead us to hypothesize that participants' evaluations may be influenced by their interaction (or lack thereof) with the GenAI model. Although all of these phenomena have their own idiosyncrasies, they all refer to effects due to experiences with or exposure to a piece of information or system. Hence, we also explore a potential \textbf{experience effect,} which posits that the experience of interacting with the GenAI model can influence participants' perceptions about AI-generated art.}

\section{Methods}

We conducted a large-scale experiment to capture laypeople's opinions concerning copyright-relevant factors (the creativity, effort, and skill necessary to create art using AI), as well as the authorship and potential copyright protection of AI-generated art. Furthermore, we studied whether people exhibit biases in their opinions by evaluating hypotheses we form based on prior work regarding potential differences between the perceptions of those who use AI models to create art and those who observe and evaluate AI-generated art. In this section, we describe our experimental design, including the experimental setting and the procedures we employ to gather our data. The study was approved by our Ethics Review Board (ERB). All data and analysis scripts are available at \repo{}. 

\subsection{Setting}
To study lay perceptions of AI-generated art, we held an online AI art exhibition. We recruited laypeople to participate in the exhibition as either creators or evaluators (or both, as explained below). Participants were told that the exhibition was juried~\cite{Whatisan69:online}, meaning that not all images would be displayed in the exhibition, such that the art would be evaluated by a jury (in this case of other crowdworkers) and that an online website would display the 10 highest-rated images, the creators of which would be awarded a monetary award. Our setting not only enabled us to capture lay perceptions of AI-generated art in relation to copyright law more generally, but also allowed us to form hypotheses regarding how its incentive structure may influence these perceptions. 

Our choice of experimental setting was motivated by a series of considerations. Our first consideration is related to the experiment's ecological validity. We mimicked a real-world setting common in the art world: a juried art exhibition in which participants can submit their art online and where the best-rated submissions (i) receive recognition by being displayed in the exhibition and (ii) receive monetary rewards (see~\cite{nytexhibition,smithexhibtion} for an equivalent scenario involving AI-generated art). The monetary rewards and recognition also serve a second purpose---incentivizing creators to put effort into creating images, as real-world creators of (GenAI or human-created) images normally would. Furthermore, the promise of monetary rewards and recognition to creators may also incentivize evaluators to take the task more seriously since they are made aware of the fact that their responses influence other participants' outcomes. 

It is important to note that our exhibition differs from many juried art exhibitions in one important aspect: a part of the jury consists of creators' peers who can also submit their art and not an (unbiased) set of professionals who do not compete in the exhibition. We opted for this design choice to study egocentric effects on people's perceptions of GenAI images and their potential copyright.

\subsection{Experimental Conditions}

In our experiments, we recruited three groups of participants. First, we recruited \emph{creators}, who had the chance to use a GenAI model to create an image for consideration at the AI art exhibition and then evaluated their own creation with respect to several variables. Second, we recruited \emph{evaluators}, who were randomly assigned to one of two separate conditions: \emph{invested evaluators} and \emph{uninvested evaluators}. \emph{Invested evaluators} used the same GenAI model to create an image for the exhibition before evaluating a subset of the submissions made by \emph{creators}. \emph{Uninvested evaluators} did not use the GenAI model and instead only evaluated \emph{creators'} images. Figure~\ref{fig:methods} presents a high-level overview of our experimental conditions. 

Before providing more details about each of the three experimental conditions to participants, we introduced the study setting, which was similar across all three treatments. On the study's landing page, we described our experiment and gathered participants' informed consent. We explained the study setting and---depending on the experimental condition to which they were assigned---informed participants that they would be generating and/or evaluating images for an AI-generated art exhibition. After reading the task description, participants were asked comprehension check questions to ensure that they understood the task; participants could not continue with the study until they answered the questions correctly. Furthermore, we informed participants that the creators of the 10 highest-rated submissions to our juried art exhibition would have their submissions displayed in an online gallery and earn a monetary reward of 25 USD. Finally, we provided participants with the following definition of GenAI models:
\begin{quote}
     Generative AI models like the one [you will use/used by other Prolific workers] learn patterns and relationships from a dataset of human-created content (e.g., human-created images like paintings and photographs). 

     When a person prompts the AI model (e.g., asks the AI model to generate an image of the sky), the AI model uses the patterns it learned from human-created content to generate the requested content (e.g., an image of the sky).
\end{quote}


\begin{figure*}[t]
\centering  
\includegraphics[width=.47\linewidth]{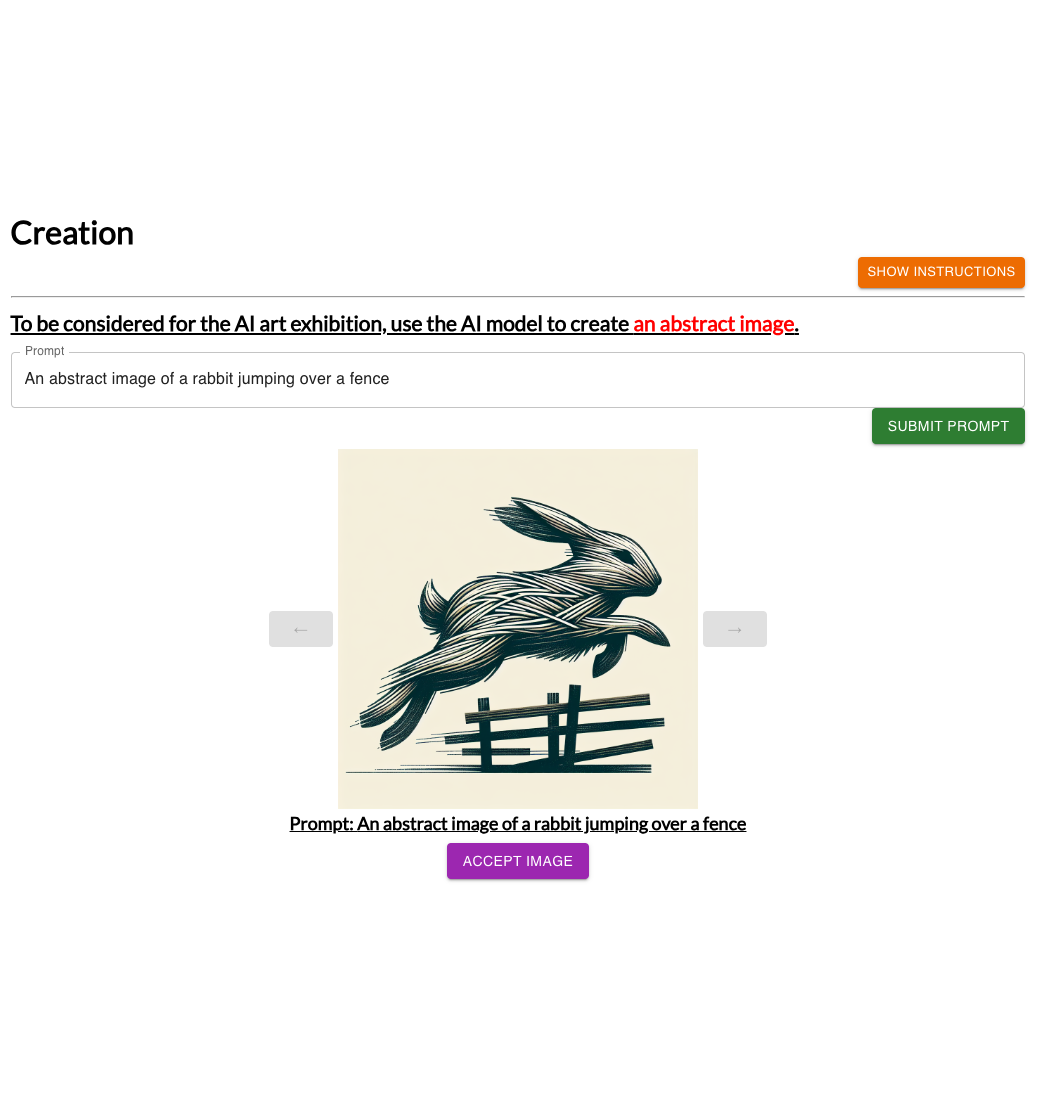}
    \hfill
\includegraphics[width=.47\linewidth]{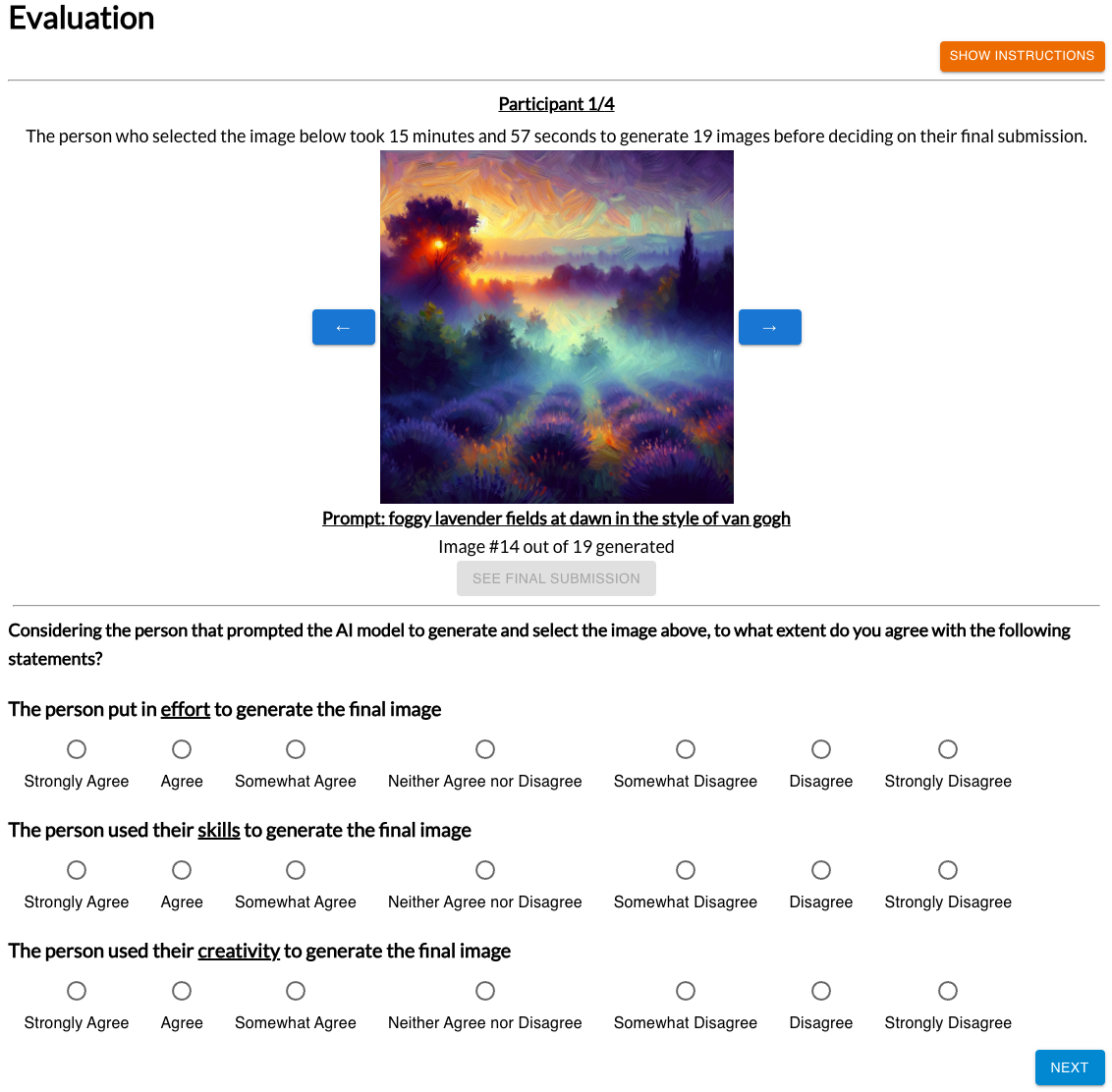}
    \caption{Example screenshots of the study's user interface.}
    \label{fig:screenshots}
    \Description{The figure is divided into two subfigures. The left subfigure shows the study's interface in which participants generated images using the AI model. The interface contains some instructions, a text box, buttons, and the generated image. The right sub-figure shows the interface used to evaluate the submitted images. The interface shows the generated images along with some metadata, followed by the survey questions.}
\end{figure*}

\subsubsection{Creators}

The experiment commenced with a three-step tutorial, in which participants were taught how to use the GenAI model to generate images. In the first step, participants were taught to write prompts. Participants were told that the AI model would generate an image according to their written instructions and were asked to instruct the AI model to generate ``an image of a cat in a comic-book style of art.'' In the second step, participants were informed that they could ask the GenAI model to generate as many images as they wanted. We emphasized that AI model they were using (DALL-E 3~\cite{betker2023improving}) did not keep previous instructions in its memory, meaning that they had to fully describe the image they intended to generate each time. As an exercise, participants were told to instruct the AI model to generate ``an image of a rabbit in an abstract style of art.'' Third, participants were shown how to navigate between all the images they created (with arrows located beside the image) and how to select the image they would like to submit for consideration at the exhibition (by clicking on a button).

After completing the tutorial, creators were asked to generate their submission to our AI art exhibition (see left panel of Figure~\ref{fig:screenshots}). They were randomly assigned the task of generating \emph{a portrait}, \emph{an image of a landscape}, or \emph{an abstract image} on a between-subjects basis to provide participants with some guidance and ensure some variance in our data. Participants could generate as many images as they wanted and select whichever image they preferred, independently of the order in which they were generated. After selecting the image they wanted to submit to the AI art exhibition, participants evaluated the image they submitted with respect to several variables, all of which we explain in Section \ref{sec:measures}.

\subsubsection{Invested Evaluators}

Similarly to creators, invested evaluators also completed a tutorial before generating, selecting, and evaluating their own submission to the AI art exhibition. After evaluating their own submission, they evaluated submissions made by four randomly selected creators, one at a time (see right panel of Figure~\ref{fig:screenshots}). When evaluating a creator's submission, participants had access not only to the image the creator submitted to the AI art exhibition, but also to all images generated by that creator and their respective prompts, as well as some descriptive statistics about the generation process (namely, the number of images that the creator generated before deciding on their final submission and how long they took to create and select it).

\subsubsection{Uninvested Evaluators}

Unlike invested evaluators, uninvested evaluators did not have the chance to complete the tutorial nor to create and submit an image to the AI art exhibition. Instead, participants only evaluated submissions made by four randomly selected creators, one at a time. They used the same interface as invested evaluators (see right panel of Figure~\ref{fig:screenshots}).

\subsection{Measures}
\label{sec:measures}

All participants evaluated their own creations and/or images created by others. Each image was judged according to four groups of questions. Each of the three first question groups addressed a separate research question (see Section~\ref{sec:intro}), whereas the last group helped determine which art would be included in the AI art exhibition.

\begin{enumerate}
     \item Factors associated with copyright decisions (RQ1):\footnote{{We did not provide a definition of what creativity, effort, and skills mean in the context of our study because of the lack of clear legal definition of these factors, as well as disagreement on whether these factors matter when deciding whether a work warrants copyright (see Section~\ref{sec:factors}). Whether a work satisfies these conditions is open to legal interpretation and argument. Hence, we decided not to bias participants' responses with respect to a particular definition of these factors and instead capture lay interpretations of these factors. Had we provided a particular definition to participants, we would have biased them to answer our questions in a particular way. In Section~\ref{sec:discussion}, we explore whether lay opinions are aligned or at odds with existing legal definitions and interpretations.}}
    \begin{enumerate}
        \item Creativity: Creativity is one of the most important factors determining whether works are eligible for copyright. For instance, under US copyright law, a work must have at least ``a modicum of creativity''~\cite{1991feist}. Similarly, EU courts have clarified that creations must be the result of the author's ``free and creative choices''~\cite{hugenholtz2021copyright}. Participants evaluated the creativity involved in the image generation process by agreeing with the following statement on a 7-point scale (0 = Strongly Disagree, 6 = Strongly Agree): ``[I/The person] used [my/their] creativity to generate the final image.''
        
        \item Effort: Although some jurisdictions have rejected that mere effort warrants copyright protection (i.e., the sweat of the brow doctrine in the US~\cite{1991feist}),
        other countries, such as Australia, suggest that effort could be sufficient~\cite{eshraghian2020human}. Scholars have also suggested that copyright decisions regarding GenAI outputs should consider the effort put in by users~\cite{guadamuz2017androids}. Perceived effort was evaluated by agreeing with the following statement on the same 7-point scale: ``[I/The person] put in effort to generate the final image.''

        \item Skills: There exists a legal precedent in Canada~\cite{2004cch} that states that copyright protection requires the exercise of non-trivial skills. Similarly, Australian law~\cite{1964ladbroke,eshraghian2020human} suggests that skill is sufficient for copyright. Participants evaluated skills by agreeing with the following statement on the same 7-point scale: ``[I/The person] used [my/their] skills to generate the final image.''
    \end{enumerate}
    
    \item Attribution of authorship (RQ2): The question of copyright is closely related to the question of authorship. One of the reasons why AI-generated art is not eligible for copyright protection in the US is that it lacks \emph{human} authorship~\cite{copyrightverge}. Hence, deciding who is the author (or authors) of a GenAI output is crucial to determining whether it is eligible for copyright protection. Although the user and the AI model itself could be seen as authors due to their role in the generation process, it is also possible that laypeople perceive other actors as authors, such as the artists whose creations were used to train the GenAI model and the company that developed it. Hence, we asked participants to what extent they agreed that these four entities are authors using the same 7-point agreement scale:
    \begin{enumerate}
        \item User: ``[I am/The person who used the AI model to generate this image is] an author of this image.''
        \item AI Model: ``The AI model itself is an author of this image.''
        \item Company: ``The company that developed the AI model is an author of this image.''
        \item Data Contributors: ``The artists whose creations were used for training the AI model are authors of this image.''
        
    \end{enumerate}
    \item Attribution of rights (RQ3): Copyright law grants several exclusive rights to the holder, such as the right to distribute, reproduce, and display the work, as well as make copies and prepare derivative materials~\cite{uscopyrightlaw}. We captured participants' opinions about two of these rights: 1) the right to display and 2) the right to make copies.
    \\
    It is important to note that it is legal to use copyrighted material under certain conditions, particularly when the work is used for non-commercial purposes (e.g., according to the US fair use doctrine). We thus collected people's opinions about the two rights in both commercial and non-commercial settings. 
    \\
    Respondents indicated whom they think should have rights over the image out of a list of entities: 1) the user, 2) the AI model, 3) the company that developed the AI model, 4) data contributors, 5) anyone, and 6) someone else (followed by an open-ended text box for indicating whom). Participants could select as many entities as they wished for each right and setting combination (e.g., right to display commercially). Entities 1-4 were described as in the authorship question presented above.
    
    \item Score evaluation: Participants also evaluated each image using an 11-point scale: ``On a scale from 0 (Very bad) to 10 (Very good), how would you evaluate this image?'' Responses to these questions determined which images were selected for the exhibition and thus received the monetary award.
\end{enumerate}

First, participants answered the groups of questions about factors associated with copyright decisions (1) and authorship (2). The order of these two groups of questions was randomized, as was the order of the questions within each group. Questions related to attribution of rights (3) followed, also in random order. Finally, participants scored the images to determine awards (4). For readability, some questions were rephrased depending on whether respondents were evaluating their own or other people's creations (see above for the exact phrasing). Before completing the study, participants also answered a series of exploratory questions, such as how often they use image and text GenAI models and demographic questions.

\subsection{Hypotheses: Egocentricity, Competition, and Experience Effects}
\label{sec:hypotheses}

{We build upon prior work studying egocentric effects (see Section~\ref{sec:biases}) and study how they may emerge in the context of GenAI. We formed two hypotheses concerning egocentric effects on lay perceptions of AI-generated images in relation to copyright. In H1, we compare the judgments of different participants about the same image. Namely, we hypothesize that the creator of an image will judge their own image more favorably than other participants. In H2, we compare the judgments of the same participant about different images. We hypothesize that participants will judge images they create more favorably than they will judge images created by others. }

\begin{itemize}[label={},leftmargin=*, itemsep=5pt]
    \item \textbf{H1)} \underline{Egocentric effect between participants:} Creators will 1) evaluate images more favorably and 2) be more likely to identify users as authors and right-holders when judging their own creations compared to invested and uninvested evaluators judging the same images. 
    \item \textbf{H2)} \underline{Egocentric effect between images:} Invested evaluators will 1) evaluate images more favorably and 2) be more likely to identify users as authors and right-holders when judging their own creations than when judging other people's images.
\end{itemize}

{We also explore the aforementioned competition effect. Creators' and invested evaluators' outcomes depended on other participants' assessments since only the highest-rated images would receive monetary rewards. Invested evaluators had the opportunity to inflate their own relative rating not only by giving high ratings to their own submission (in line with H2), but also by giving low scores to other people's submissions. That is, they could benefit from sabotaging their competition. In contrast, uninvested evaluators had no incentive to sabotage other participants since they were not competing for the monetary rewards. Therefore, we hypothesize that the incentive structure associated with our juried art exhibition may lead those with something at stake (invested evaluators) to judge others' submissions more harshly than those who do not have anything to gain from doing so (uninvested evaluators).}

\begin{itemize}[label={},leftmargin=*, itemsep=5pt]
    \item \textbf{H3)} \underline{Competition effect:} Invested evaluators will 1) evaluate other people's images less favorably and 2) be less likely to identify other users as authors and right-holders than uninvested evaluators.
\end{itemize}

{Finally, we also make hypotheses based on the experience effect. Uninvested evaluators did not have an opportunity to interact with the GenAI model and to create AI-generated art in our experiments. On the other hand, creators and invested evaluators repeatedly interacted with the model throughout the tutorial and creation task, observing how changing their inputs influences the model's outputs and learning how to use it to create AI-generated art. Consequently, we hypothesize that the experience of interacting with the GenAI model will influence participants' perceptions of AI-generated art.}

\begin{itemize}[label={},leftmargin=*, itemsep=5pt]
    \item \textbf{H4)} \underline{Experience effect:} Participants who interacted with the GenAI model (i.e., creators and invested evaluators) will 1) evaluate images and 2) attribute authorship and rights differently than those who did not use the GenAI model (i.e., uninvested evaluator).
\end{itemize}

Finally, we emphasize that while we gather data on how participants attribute authorship and rights to various entities, our hypotheses focus on attributions of authorship and rights to one specific entity: \emph{users}. We investigate the remaining entities exploratively.

\subsection{Analysis Plan}
\label{sec:analysis_plan}

We used regressions to analyze our data. We treated participants' assessments of factors, authorship, and score evaluations as continuous dependent variables in linear regressions. Participants' attributions of rights were treated as binary dependent variables and modeled using logit regressions. To account for repeated measurements across participants and images (i.e., participants evaluated several images, and images were evaluated by several participants), we initially planned to use mixed-effects linear regressions with crossed random intercepts for images and participants. However, due to convergence issues, we instead opted for regressions with two-way clustered standard errors~\cite{correia2016feasible}. 

Our primary independent variable is a dummy variable encoding both the i) treatment condition to which the participants were assigned and ii) whether the data point refers to an evaluation of their own image or someone else's creation. Hence, our dummy variable has four levels representing 1) \emph{creators} judging their \emph{own} images; 2) \emph{invested evaluators} rating their \emph{own} creations; 3) \emph{invested evaluators} assessing \emph{others'} images; and 4) \emph{uninvested evaluators} judging \emph{others'} submissions.

We tested for differences between pairs of treatments by estimating their contrasts (i.e., by estimating the difference between the treatments' estimated regression coefficients) and applied Bonferroni corrections to account for multiple comparisons. We conducted such pairwise comparisons only on pairs of treatments that are the subject of the hypotheses described above. In Section~\ref{sec:results}, we discuss all pairwise differences that were significant at the $\alpha$ < .05 level.

Finally, as a robustness check, we repeated the analysis described above while accounting for three different sets of covariates by including them as additional independent variables:

\begin{enumerate}
    \item {Participant-level variables: we included ten variables concerning the evaluator: their self-reported 1) gender, 2) age, and 3) race; how often they use GenAI models for generating 4) images and 5) text; whether they had already participated in a study in which they were asked to 6) generate and 7) evaluate images; and whether they have any training in professions related to 8) art, 9) computer science, or 10) law.}
    \item Image-level variables: we account for the 1) number of images that creators generated before selecting one for the exhibition; 2) how long the process took; 3) the length of the selected prompt; and 4) the type of image they were asked to generate (portrait, landscape, or abstract).
    \item Order variable: we also included a variable indicating the order in which the image was shown to an evaluator. For instance, if a measurement refers to the second image that participants evaluated, this variable is equal to two. This analysis not only provides robustness to our results but also explores any order effects in participants' evaluations.  
\end{enumerate}

{Our results are robust to all three groups of covariates. That is, pairwise comparisons between treatments remain qualitatively and quantitatively similar upon including any of the three groups of covariates, barring minor changes in the significance of borderline results. For simplicity and brevity, we do not discuss the results of models with participant- and image-level covariates in the paper. However, we discuss order effects when the coefficient is significant at the $\alpha$ < .05 level. We present all regressions---with and without covariates---in the Appendix and make all of our data available for replication.}

\subsection{Data Collection \& Participants}

For the main study, we recruited 450 participants on Prolific~\cite{palan2018prolific}. First, we recruited 100 participants to complete the study as creators, followed by an additional 350 participants divided equally between invested and uninvested evaluators. We targeted US residents who were fluent in English and had completed at least 50 tasks on Prolific with an approval rate of over 95\%. Participants were sampled at different hours over several days to mitigate sampling biases that may occur due to time~\cite{casey2017intertemporal}.

{Due to technical problems, responses from two participants had to be dropped because they were not saved completely. We discarded responses from 16 participants who failed any of two instructed response questions. Finally, we discarded judgments made by (invested and uninvested) evaluators regarding images generated by creators that were removed due to attention check failures. Our final sample comprises 432 participants, out of which 95 were creators, 169 were invested evaluators, and 168 were uninvested evaluators. All participants were paid \$3.50 USD for their participation regardless of their treatment condition to keep monetary incentives constant (approximately \$12.00 USD per hour).}

{Nearly half (49.77\%) of participants identified themselves as women, 46.99\% as men, and 3.23\% as non-binary or chose not to disclose their gender. Participants' mean age was 40.3 years old ($SD$ = 14.0), with the youngest respondent being 20 years old and the oldest 81 years old. Our sample is slightly more diverse than the US population in terms of race, with 10.19\% of participants describing themselves as Asian, 15.05\% as Black or African American, and 61.34\% as White. Only 7.41\% of participants reported having prior training in professions related to law, while 23.61\% stated that they had training in computer science-related professions and 26.39\% in art-related occupations.}

In addition to the participants recruited for the main study, we recruited an additional sample of evaluators to determine which art should be included in the exhibition. In the main study, only creators' submissions received evaluations from other participants. Specifically, each of the creators' submissions was evaluated by approximately 13 other participants from the pool of invested and uninvested evaluators. On the other hand, invested evaluators' submissions were only evaluated by themselves. Since both creators and invested evaluators were eligible to receive awards, we recruited 115 additional participants to rate 20 of the invested evaluators' submissions each, resulting in approximately 13 evaluations for each submission to the art show. Participants in this additional study only evaluated images with respect to quality to determine awards. These participants were paid 1.60 USD (approximately \$12.00 USD per hour).

\section{Results}
\label{sec:results}
\begin{figure*}[!htbp]
    \centering
    \includegraphics[width=0.18\textwidth]{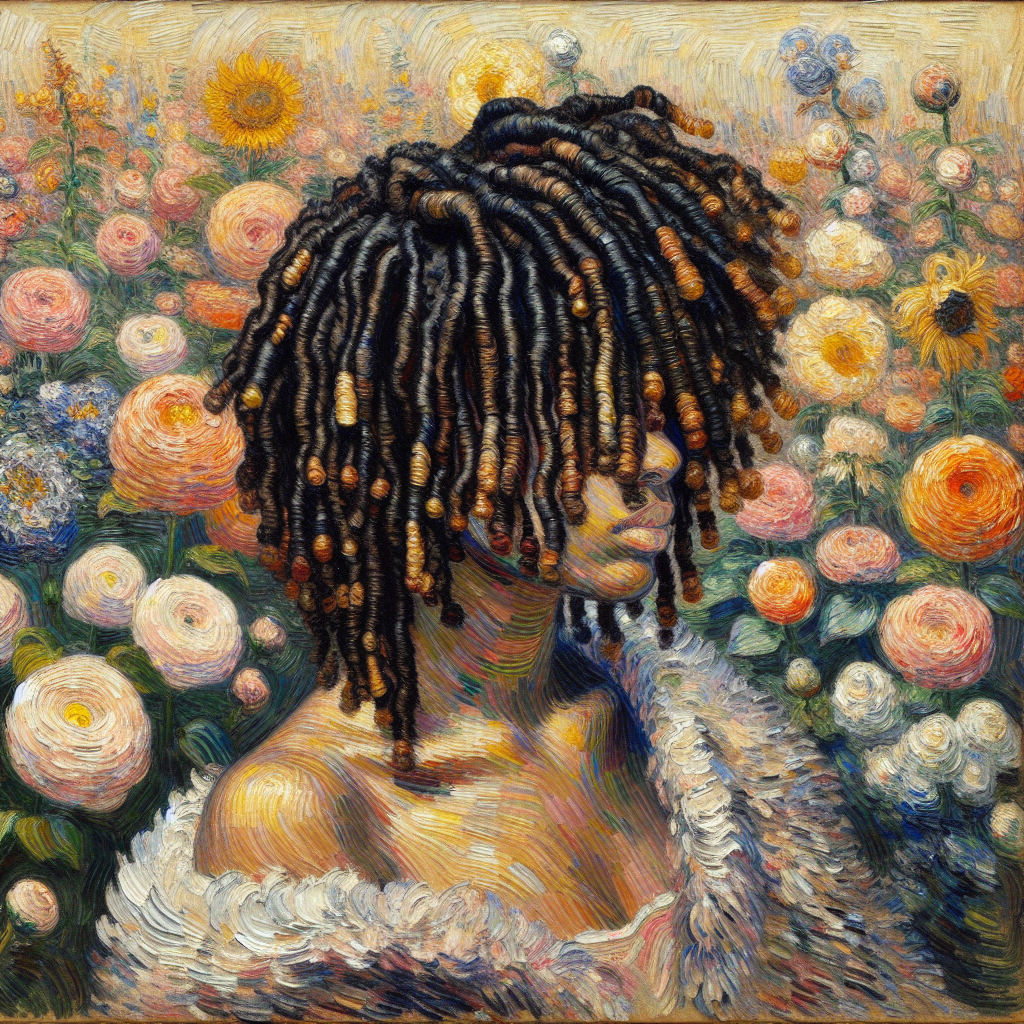}
    \includegraphics[width=0.18\textwidth]{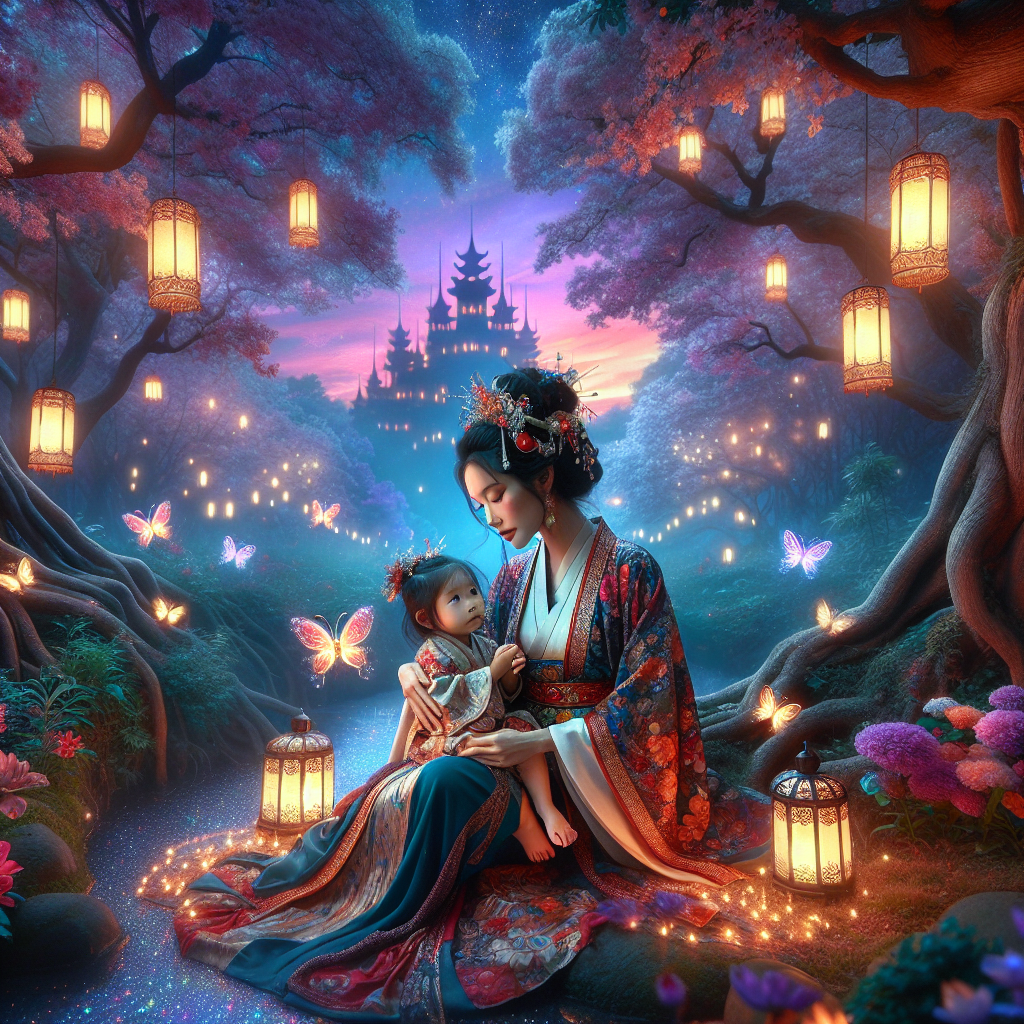}
    \includegraphics[width=0.18\textwidth]{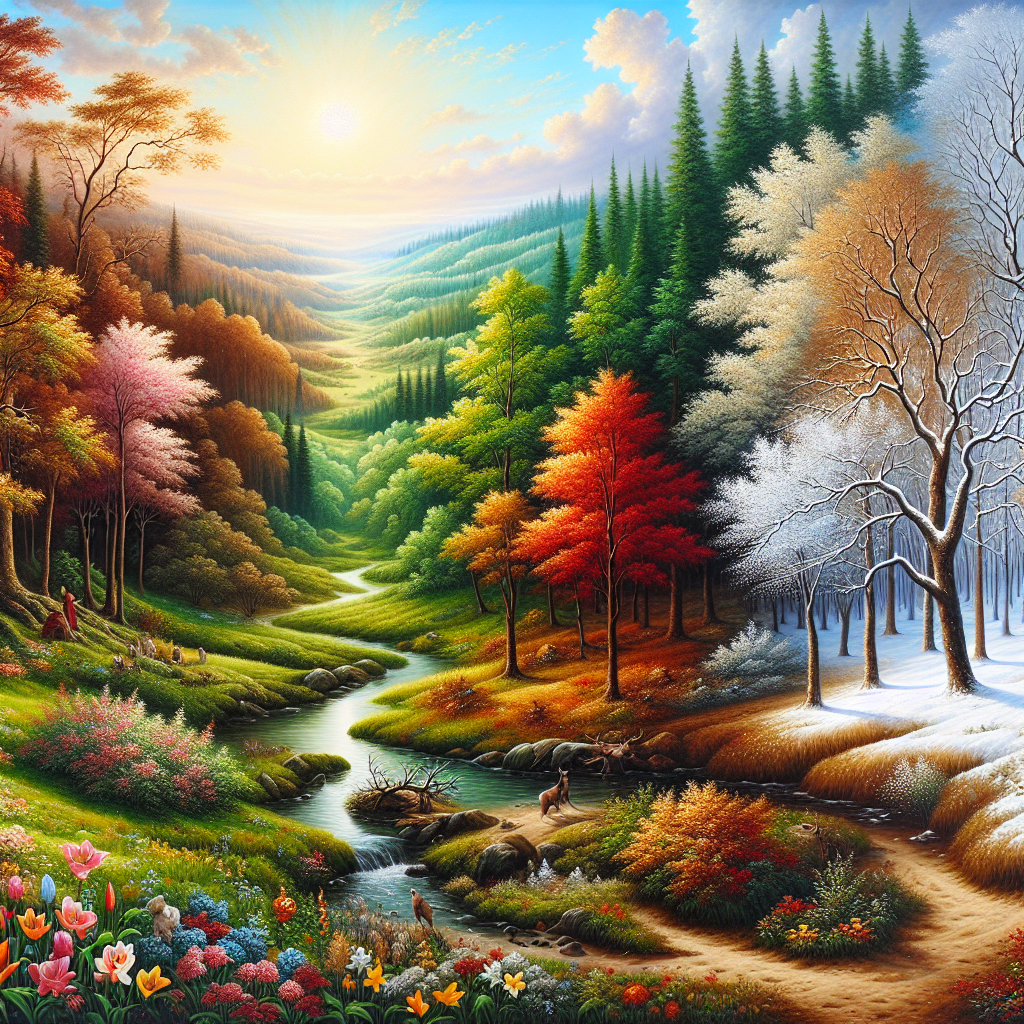}
    \\
    \includegraphics[width=0.18\textwidth]{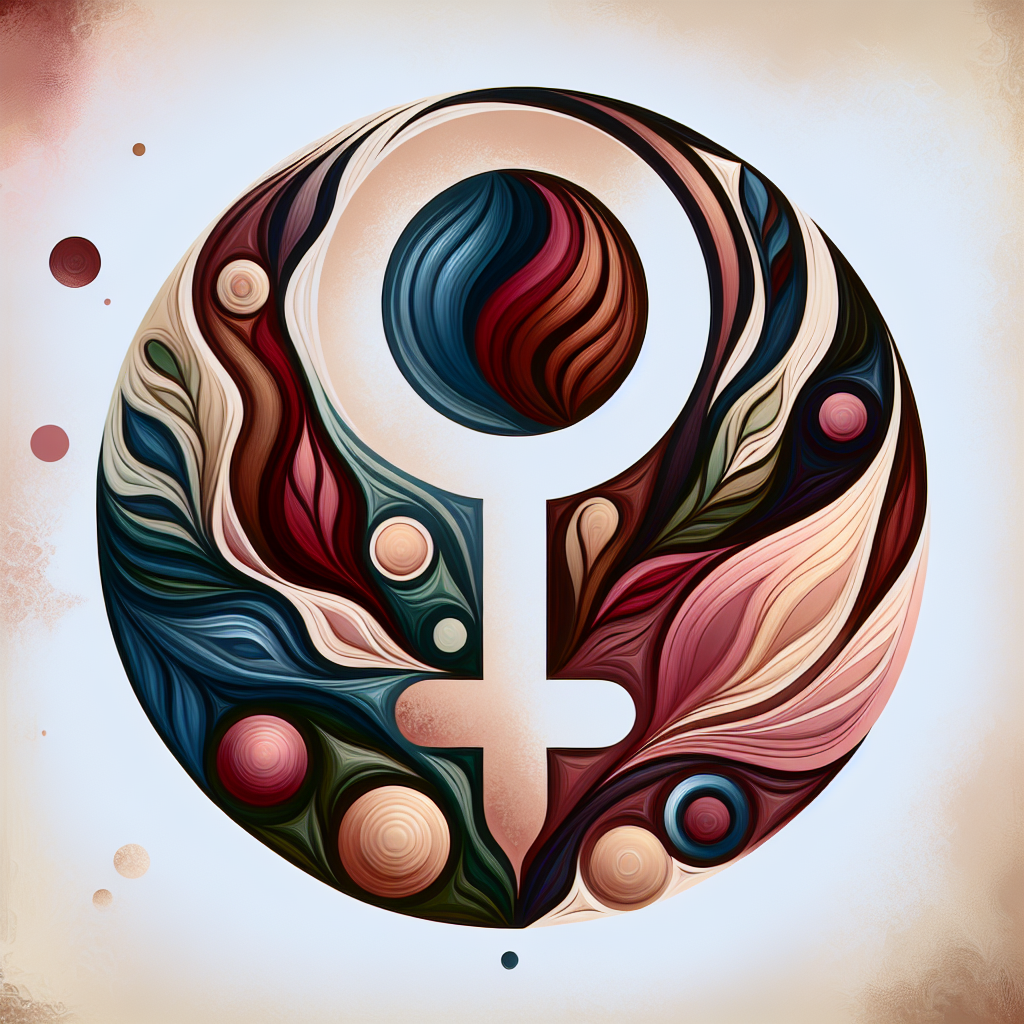}
    \includegraphics[width=0.18\textwidth]{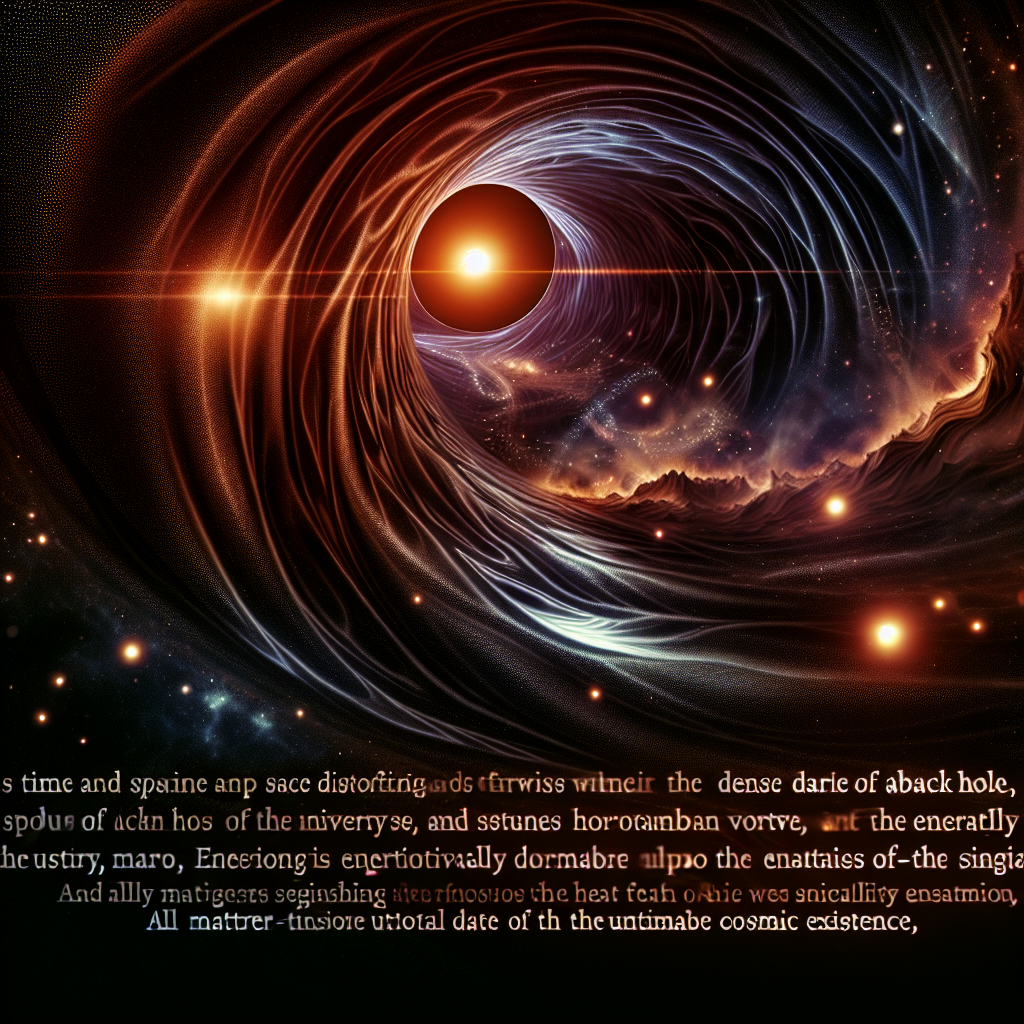}
    \includegraphics[width=0.18\textwidth]{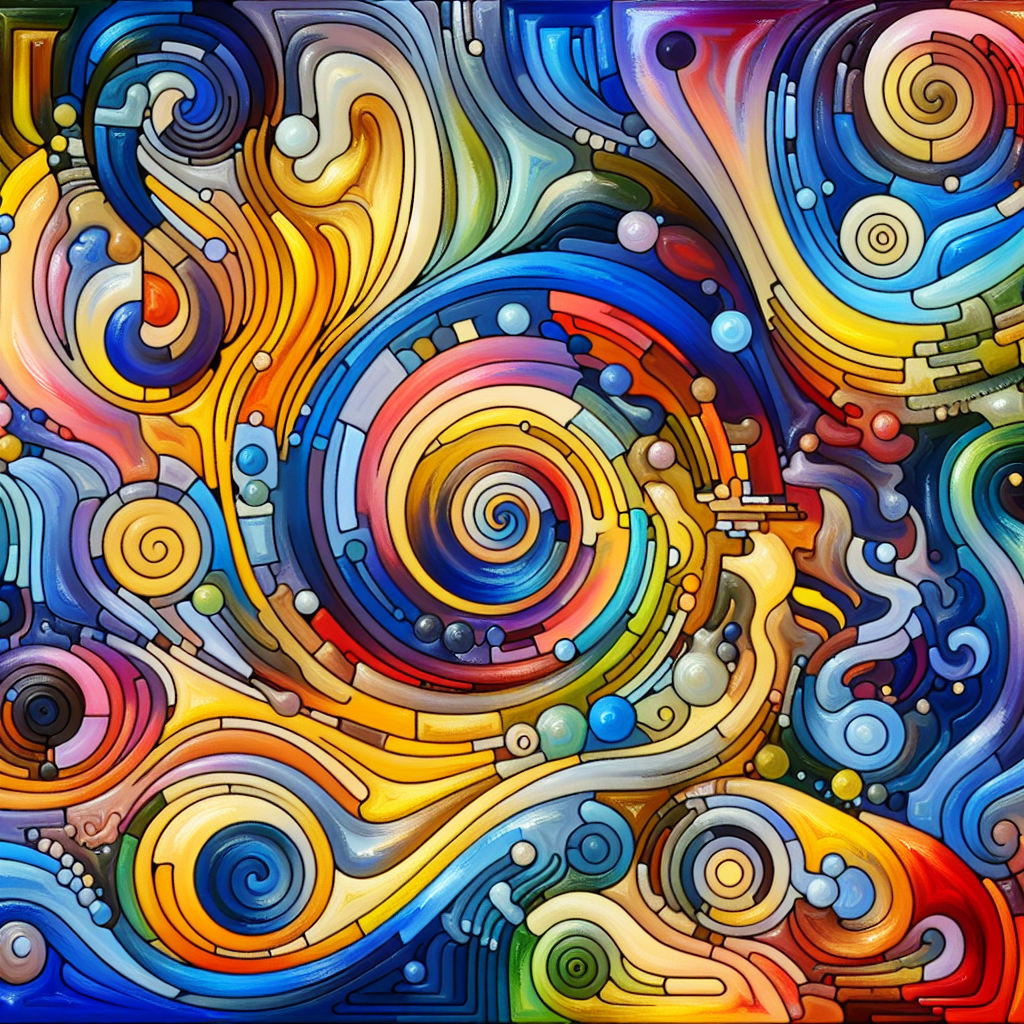}
    \caption{Example images generated by creators for our AI art exhibition. The images in the top row were among the best evaluated, while those in the bottom row were judged poorly.}
        \label{fig:exampleimages}
    \Description{Six example images. The top-left image depicts a person with dark skin and braids in front of a flower field. The top-center image depicts an Asian woman holding a baby in a fairytale-like setting. The top-right image portrays a forest. The bottom-left image is an abstract picture with the symbol of the female gender symbol in the center. The bottom-center image depicts a dark universe with some stars. The bottom-right image is a colorful abstract image with circular patterns.}
\end{figure*}

A total of 264 participants submitted an image to the AI art exhibition (95 creators and 169 invested evaluators). These participants generated a median of two images (mean = 4.35, $SD$ = 5.10). Although 90 of these participants generated only one image, 31 of them created 10 or more. Creators generated a median of 2 images (mean = 5.01, $SD$ = 6.40), while invested evaluators created a median of 3 images (mean = 3.98, $SD$ = 4.16). Participants took an average of 5.87 minutes ($SD$ = 5.86) to create and select an image for the exhibition. Invested evaluators (mean = 5.95 minutes, $SD$ = 5.61) spent slightly more time generating images than creators (mean = 5.73, $SD$ = 6.23). Figure~\ref{fig:exampleimages} presents some example images.  

\subsection{RQ1: Perceived Creativity, Effort, and Skills}

\begin{figure*}[!htbp]
    \centering
    \includegraphics[width=\textwidth]{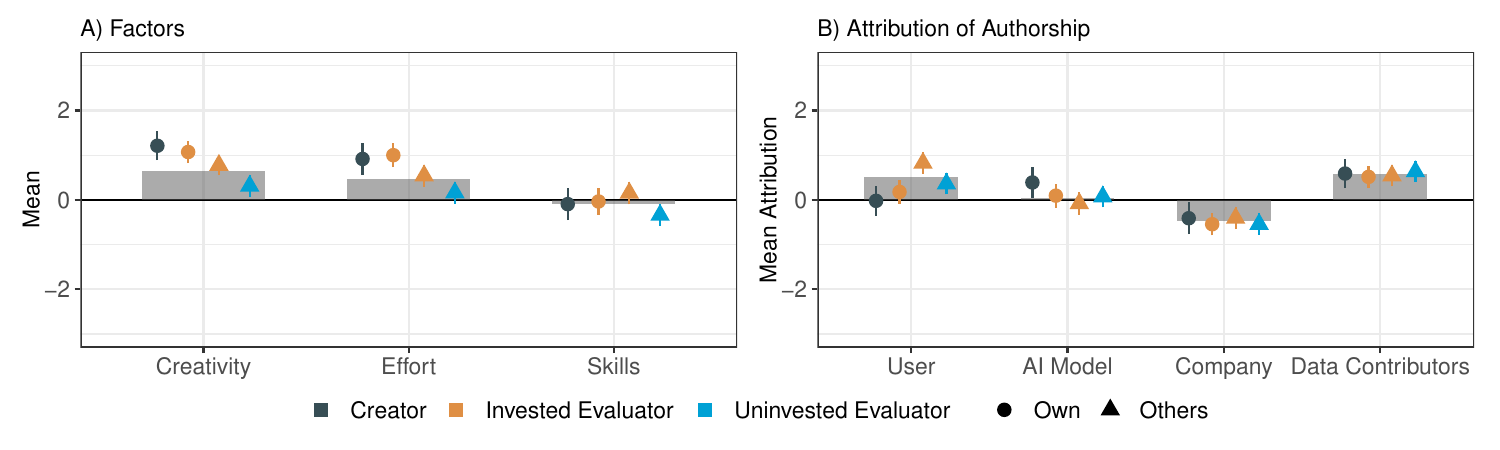}
    \caption{A) Perceived creativity, effort, and skills involved in generating images using GenAI for our AI art exhibition. B) Perceived authorship of the user, the AI model, the company that developed the AI model, and data contributors (i.e., those whose creations were used to train the AI model). Gray bars present the mean value across all conditions, while circles and triangles represent mean values in each treatment condition according to the legend. Error bars correspond to 95\% confidence intervals.}
    \label{fig:factors_authorship}
    \Description{The figure consists of two panels. Panel A (on the left) is a bar graph representing the mean values of perceived creativity, effort, and skills. The plot also includes mean values depending on the treatment condition. Panel B (on the right) is an equivalent bar graph that illustrates mean attributions of authorship to the user, the AI model, the company that developed the AI model, and data contributors. The plot also shows mean values depending on the treatment condition.}
\end{figure*}

\begin{table*}[!htbp] \centering 
\small
\begin{tabular}{@{\extracolsep{5pt}} l|rrlr} 
\\[-1.8ex]\hline 
\hline \\[-1.8ex] 
Contrast & diff & SE & t-test & p-value \\ 
\hline \\[-1.8ex] 
\textbf{Creativity} & & & & \\
Creators (Own) - Uninvested Evaluators (Others) - H1 and H4 & $0.900$ & $0.191$ & t(1534) = 4.704 & \bm{$<0.001$} \\ 
Creators (Own) - Invested Evaluators (Others) - H1 & $0.434$ & $0.178$ & t(1534) = 2.431 & $0.076$ \\ 
Invested Evaluators (Own) - Invested Evaluators (Others) - H2 & $0.294$ & $0.135$ & t(1534) = 2.179 & $0.147$ \\ 
Uninvested Evaluators (Others) - Invested Evaluators (Others) - H3 and H4 & $-0.466$ & $0.145$ & t(1534) = -3.226 & \bm{$0.006$} \\ 
Invested Evaluators (Own) - Uninvested Evaluators (Others) -  H4 & $0.761$ & $0.179$ & t(1534) = 4.250 & \bm{$<0.001$} \\ 
\\[-1.8ex] 
\textbf{Effort} & & & & \\
Creators (Own) - Uninvested Evaluators (Others) - H1 and H4 & $0.754$ & $0.193$ & t(1534) = 3.899 & \bm{$0.001$} \\ 
Creators (Own) - Invested Evaluators (Others) - H1 & $0.373$ & $0.206$ & t(1534) = 1.809 & $0.354$ \\ 
Invested Evaluators (Own) - Invested Evaluators (Others) - H2 & $0.458$ & $0.141$ & t(1534) = 3.252 & \bm{$0.006$} \\ 
Uninvested Evaluators (Others) - Invested Evaluators (Others) - H3 and H4 & $-0.381$ & $0.149$ & t(1534) = -2.556 & $0.053$ \\ 
Invested Evaluators (Own) - Uninvested Evaluators (Others) -  H4 & $0.839$ & $0.186$ & t(1534) = 4.512 & \bm{$<0.001$} \\ 
\\[-1.8ex] 
\textbf{Skills} & & & & \\
Creators (Own) - Uninvested Evaluators (Others) - H1 and H4 & $0.245$ & $0.212$ & t(1534) = 1.159 & $1.000$ \\ 
Creators (Own) - Invested Evaluators (Others) - H1 & $-0.249$ & $0.227$ & t(1534) = -1.097 & $1.000$ \\ 
Invested Evaluators (Own) - Invested Evaluators (Others) - H2 & $-0.190$ & $0.119$ & t(1534) = -1.592 & $0.558$ \\ 
Uninvested Evaluators (Others) - Invested Evaluators (Others) - H3 and H4 & $-0.494$ & $0.162$ & t(1534) = -3.055 & \bm{$0.011$} \\ 
Invested Evaluators (Own) - Uninvested Evaluators (Others) -  H4 & $0.305$ & $0.193$ & t(1534) = 1.576 & $0.576$ \\  
\hline
\end{tabular}
  \caption{Pairwise comparisons of perceived creativity, effort, and skills between treatment conditions. We only test the contrasts relevant to our hypotheses presented in Section~\ref{sec:hypotheses}. We apply Bonferroni corrections to account for multiple comparisons.} 
  \label{tab:factor_pairwise} 
\end{table*} 

Figure~\ref{fig:factors_authorship}A shows participants' mean evaluations regarding the creativity, effort, and skills involved in generating images with GenAI. Participants somewhat agreed that creativity (\msd{0.642}{1.73}) and effort (\msd{0.458}{1.81}) were necessary to create the images. In contrast, judgments concerning skills were closer to zero (\msd{-0.087}{1.82}), meaning that, on average, participants neither agreed nor disagreed that creators used their skills to generate images with GenAI. Table~\ref{tab:factor_pairwise} presents pairwise comparisons of perceived creativity, effort, and skills between treatment conditions. Below, we discuss the results of pairwise comparisons for each factor separately.

\subsubsection{Creativity}
Concerning creativity, uninvested evaluators rated images lower than creators (\diff{0.900}) and invested evaluators, both when the latter evaluated their own creations (\diff{0.761}) and other people's images (\diff{-0.466}). That is, our results regarding creativity support H4, and they partially support H1. We also found an effect in the opposite direction than the one hypothesized in H3---invested evaluators assigned higher creativity scores to creators than uninvested evaluators did.

\subsubsection{Effort}
Our analysis shows that uninvested evaluators assigned lower effort ratings than creators (\diff{0.754}) and invested evaluators judging their own images (\diff{0.839}), offering partial support to H1 and H4. We also find that invested evaluators assigned higher effort ratings to their own creations than to others' (\diff{0.458}), in line with H2.

\subsubsection{Skills}
The only significant difference in judgments of skills was found between uninvested and invested evaluators when the latter evaluated other people's images (\diff{-0.494}). Invested evaluators attributed more skills to creators than their uninvested counterparts. That is, as for creativity, we find an effect in the opposite direction than the one hypothesized in H3.

\subsection{RQ2: Perceived Authorship}

\begin{table*}[!htbp] \centering 
\small
\begin{tabular}{@{\extracolsep{5pt}} l|rrlr} 
\\[-1.8ex]\hline 
\hline \\[-1.8ex] 
Contrast & diff & SE & t-test & p-value \\ 
\hline \\[-1.8ex] 
\textbf{User} & & & & \\
Creators (Own) - Uninvested Evaluators (Others) - H1 and H4 & $-0.382$ & $0.206$ & t(1534) = -1.855 & $0.319$ \\ 
Creators (Own) - Invested Evaluators (Others) - H1 & $-0.845$ & $0.210$ & t(1534) = -4.018 & \bm{$<0.001$} \\ 
Invested Evaluators (Own) - Invested Evaluators (Others) - H2 & $-0.646$ & $0.103$ & t(1534) = -6.300 & \bm{$<0.001$} \\ 
Uninvested Evaluators (Others) - Invested Evaluators (Others) - H3 and H4 & $-0.463$ & $0.170$ & t(1534) = -2.728 & \bm{$0.032$} \\ 
Invested Evaluators (Own) - Uninvested Evaluators (Others) -  H4 & $-0.183$ & $0.180$ & t(1534) = -1.017 & $1.000$ \\ 
\\[-1.8ex] 
\textbf{AI Model} & & & & \\
Creators (Own) - Uninvested Evaluators (Others) - H1 and H4 & $0.319$ & $0.221$ & t(1534) = 1.442 & $0.747$ \\ 
Creators (Own) - Invested Evaluators (Others) - H1 & $0.470$ & $0.220$ & t(1534) = 2.132 & $0.166$ \\ 
Invested Evaluators (Own) - Invested Evaluators (Others) - H2 & $0.175$ & $0.098$ & t(1534) = 1.788 & $0.370$ \\ 
Uninvested Evaluators (Others) - Invested Evaluators (Others) - H3 and H4 & $0.151$ & $0.178$ & t(1534) = 0.846 & $1.000$ \\ 
Invested Evaluators (Own) - Uninvested Evaluators (Others) -  H4 & $0.024$ & $0.182$ & t(1534) = 0.133 & $1.000$ \\ 
\\[-1.8ex] 
\textbf{Company} & & & & \\
Creators (Own) - Uninvested Evaluators (Others) - H1 and H4 & $0.132$ & $0.213$ & t(1534) = 0.619 & $1.000$ \\ 
Creators (Own) - Invested Evaluators (Others) - H1 & $-0.003$ & $0.217$ & t(1534) = -0.015 & $1.000$ \\ 
Invested Evaluators (Own) - Invested Evaluators (Others) - H2 & $-0.137$ & $0.088$ & t(1534) = -1.557 & $0.598$ \\ 
Uninvested Evaluators (Others) - Invested Evaluators (Others) - H3 and H4 & $-0.135$ & $0.177$ & t(1534) = -0.764 & $1.000$ \\ 
Invested Evaluators (Own) - Uninvested Evaluators (Others) -  H4 & $-0.002$ & $0.176$ & t(1534) = -0.012 & $1.000$ \\
\\[-1.8ex] 
\textbf{Data Contributors} & & & & \\
Creators (Own) - Uninvested Evaluators (Others) - H1 and H4 & $-0.044$ & $0.210$ & t(1534) = -0.208 & $1.000$ \\ 
Creators (Own) - Invested Evaluators (Others) - H1 & $0.049$ & $0.204$ & t(1534) = 0.238 & $1.000$ \\ 
Invested Evaluators (Own) - Invested Evaluators (Others) - H2 & $-0.026$ & $0.096$ & t(1534) = -0.271 & $1.000$ \\ 
Uninvested Evaluators (Others) - Invested Evaluators (Others) - H3 and H4 & $0.092$ & $0.170$ & t(1534) = 0.542 & $1.000$ \\ 
Invested Evaluators (Own) - Uninvested Evaluators (Others) -  H4 & $-0.118$ & $0.173$ & t(1534) = -0.685 & $1.000$ \\ 
\hline
\end{tabular} 
  \caption{Pairwise comparisons of perceived authorship between treatment conditions. We only test the contrasts relevant to our hypotheses presented in Section~\ref{sec:hypotheses}. We apply Bonferroni corrections to account for multiple comparisons.} 
  \label{tab:author_pairwise} 
\end{table*} 

Figure~\ref{fig:factors_authorship}B shows how participants attributed authorship between the user, the AI model, the company that developed the AI model, and data contributors. Participants somewhat agreed that users (\msd{0.508}{1.71}) and data contributors (\msd{0.579}{1.70}) are authors of AI-generated images. Judgments concerning the AI model were more uncertain (\msd{0.031}{1.79}), with evaluators, on average, neither agreeing nor disagreeing that the AI model itself is an author. The company that developed the AI model had the lowest perceived authorship (\msd{-0.479}{1.72}). Table~\ref{tab:author_pairwise} presents pairwise comparisons of perceived authorship between treatment conditions. We discuss the results for each entity separately.

\subsubsection{User}

When judging others' submissions, invested evaluators attributed more authorship to creators than creators attributed to themselves (\diff{-0.845}). This effect is the opposite of the one hypothesized in H1. Moreover, invested evaluators attributed more authorship to others than to themselves (\diff{-0.646}). This effect goes against our hypothesis H2. {Finally, we found that invested evaluators attributed more authorship to creators than uninvested evaluators (\diff{-0.463}), an effect in the opposite direction to that hypothesized in H3 and partially in line with H4.}

{We also observed borderline significant order effects in participants' attribution of authorship to the user. The more images a participant evaluated, the more authorship they attributed to users ($b$ = 0.041, $SE$ = 0.022, \ttestno{1533}{1.897}{.058}). We note that the results described above are robust to including the order variable as a covariate. Invested evaluators attributed more authorship to creators (i.e., users) than creators themselves (\diff{-0.742}). Invested evaluators attributed more authorship to others than to themselves (\diff{-0.543}). However, the statistically significant difference in how invested and uninvested evaluators attributed authorship to creators disappeared when accounting for order effects ($p$ = 0.068).}

\subsubsection{Other Entities}

There were no significant differences in the perceived authorship of the AI model, the developer, and data contributors across treatments. {Nonetheless, we note a significant order effect on perceived authorship of the AI model, such that the more images participants evaluated, the less authorship they attributed to the AI model ($b$ = -0.064, $SE$ = 0.028, \ttest{1533}{-2.286}{0.05}).}

\subsection{RQ3: Attribution of Rights}

\begin{figure*}[!htbp]
    \centering
    \includegraphics[width=\textwidth]{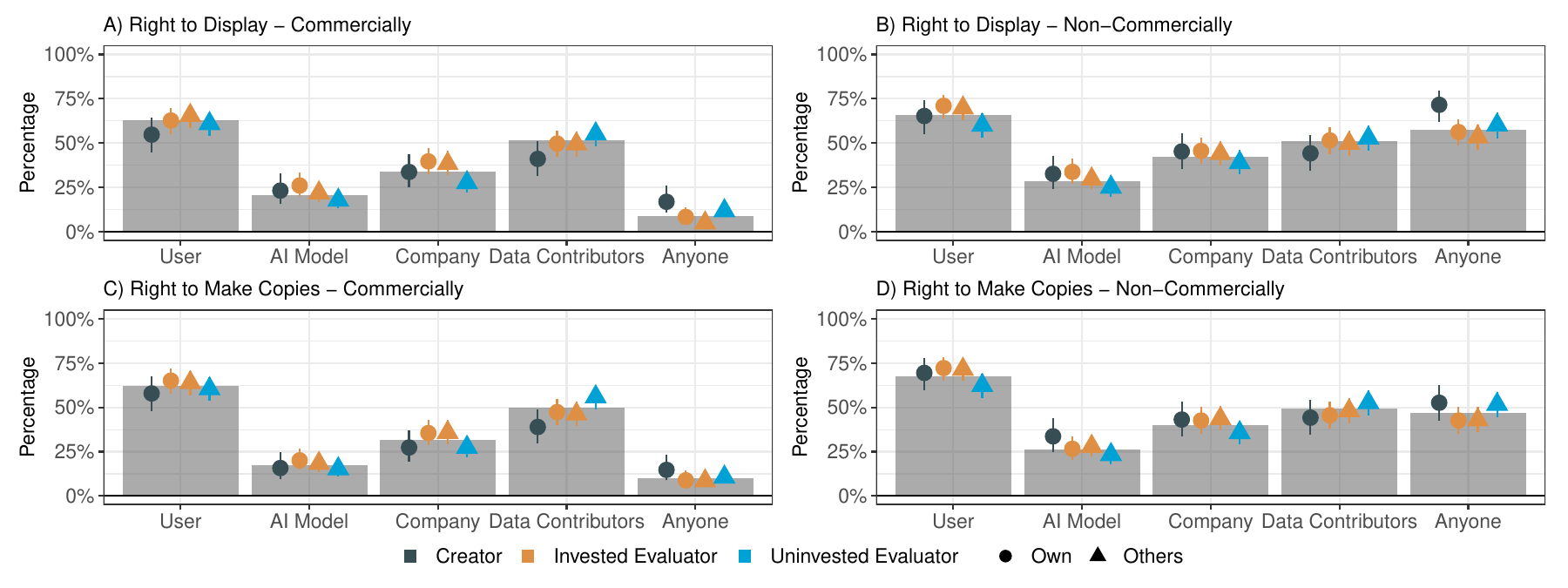}
    \caption{Percentage of participants who chose the user, the AI model, the company that developed the AI model, data contributors (i.e., those whose creations were used to train the AI model), and anyone as right-holders. We present results separately for the rights to A) display commercially, B) to display non-commercially, C) to make copies commercially, and D) to make copies non-commercially. Gray bars present the mean percentage across all conditions, while circles and triangles represent mean percentages in each treatment condition according to the legend. Error bars correspond to 95\% confidence intervals.}
    \Description{The figure consists of four panels. Each panel shows bar plots that represent the percentage of participants who picked a particular actor as the holder of a right. Actors are represented in the x-axis. The top-left panel corresponds to the right to display images commercially, the top-right panel refers to the right to display non-commercially, the bottom-left panel corresponds to the right to make copies commercially, and the bottom-right refers to the right to make copies non-commercially. Each panel also shows the data separated by experimental condition.}
    \label{fig:copyright}
\end{figure*}

Figure~\ref{fig:copyright} presents participants' opinions regarding who should hold the rights to display and make copies of AI-generated images. Users (i.e., those who used the GenAI to generate images) were selected by more than 60\% of participants for all rights, in both commercial and non-commercial settings. Data contributors were identified as rights-holders by approximately 50\% of participants across all rights and settings. The company that developed the AI model was granted the rights to display and make copies of AI-generated images by around 41\% of participants in non-commercial settings; however, only 33\% of respondents believed the company should have commercial rights over the image. The AI model was recognized as a rights-holder by around 19\% and 27\% of participants for commercial and non-commercial uses, respectively.

Participants were also able to indicate that no one should have \emph{exclusive} rights over the images by selecting that ``anyone'' should be able to display and make copies of AI-generated images. We observed clear differences in responses between commercial and non-commercial rights. Participants were more likely to support non-exclusive rights in non-commercial settings (approximately 58\% and 47\% in favor of anyone having the right to display and make copies, respectively). In contrast, when evaluating commercial rights, only a few respondents indicated that anyone should have them (approximately 9\%).

Unlike for attributions of authorship, our treatments had little effect on attributions of rights. That is, the patterns described above are consistent across all treatments, with a few exceptions detailed below. Hence, for brevity, we report the full pairwise comparison tables in the Appendix and report all of the differences identified as statistically significant directly in the text below.

We did not find support for any of our hypotheses regarding attribution of rights to users. There were only a few significant pairwise differences between conditions. Creators were relatively more likely to support non-exclusive rights to display their own creations than invested evaluators evaluating the same images (\odds{3.953}{3.330}{.005} for commercial use, \odds{2.193}{2.941}{.05} for non-commercial use). 
Finally, creators were borderline less supportive of granting the right to make copies commercially to data contributors than uninvested evaluators (\odds{0.502}{-2.589}{0.05}). No other differences were statistically significant.

\subsection{Score Evaluation}

\begin{figure*}[!htbp]
    \centering
    \includegraphics[width=.6\textwidth]{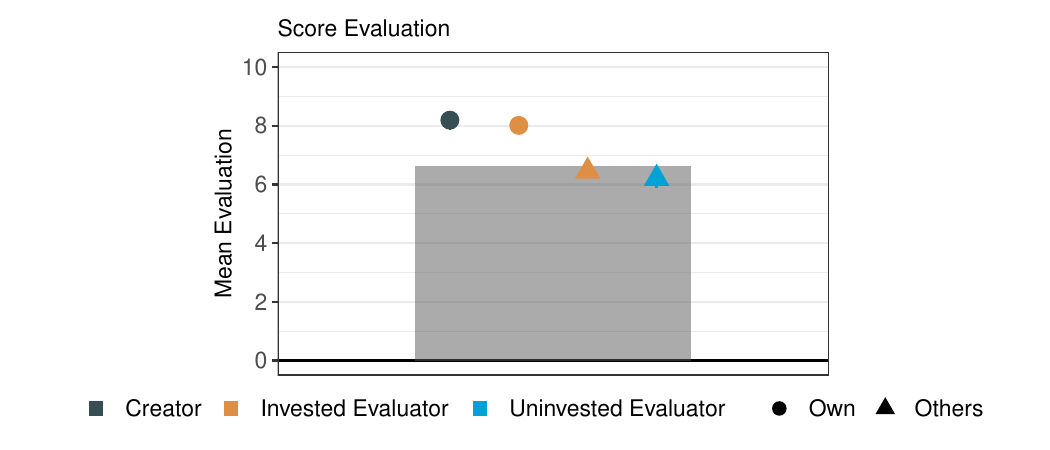}
    \caption{Score evaluations of images generated by creators. The gray bar presents the mean score across all conditions, while circles and triangles represent mean scores in each treatment condition according to the legend. Error bars correspond to 95\% confidence intervals.}
    \label{fig:eval}
    \Description{The figure depicts a bar plot that illustrates the mean score evaluation across all images in our study. The plot also shows mean values depending on the treatment condition.}
\end{figure*}

\begin{table*}[t!] \centering 
\small
\begin{tabular}{@{\extracolsep{5pt}} l|rrlr} 
\\[-1.8ex]\hline 
\hline \\[-1.8ex] 
Contrast & diff & SE & t-test & p-value \\ 
\hline \\[-1.8ex] 
Creators (Own) - Uninvested Evaluators (Others) - H1 and H4 & $1.986$ & $0.225$ & t(1534) = 8.821 & \bm{$<0.001$} \\ 
Creators (Own) - Invested Evaluators (Others) - H1 & $1.743$ & $0.213$ & t(1534) = 8.189 & \bm{$<0.001$} \\ 
Invested Evaluators (Own) - Invested Evaluators (Others) - H2 & $1.565$ & $0.190$ & t(1534) = 8.232 & \bm{$<0.001$} \\ 
Uninvested Evaluators (Others) - Invested Evaluators (Others) - H3 and H4 & $-0.243$ & $0.196$ & t(1534) = -1.242 & $1.000$ \\ 
Invested Evaluators (Own) - Uninvested Evaluators (Others) -  H4 & $1.808$ & $0.225$ & t(1534) = 8.037 & \bm{$<0.001$} \\
\hline 
\end{tabular} 
  \caption{Pairwise comparisons of image score evaluations between treatment conditions. We only test the contrasts relevant to our hypotheses presented in Section~\ref{sec:hypotheses}. We apply Bonferroni corrections to account for multiple comparisons.} 
  \label{tab:eval_pairwise} 
\end{table*} 

Figure~\ref{fig:eval} presents participants' mean score evaluations on an 11-point scale (0 = very bad, 10 = very good). These scores were used to determine who was granted the monetary award and which images were displayed in our AI art exhibition. On average, participants evaluated images slightly positively (\msd{6.62}{2.40}). Table~\ref{tab:eval_pairwise} presents pairwise comparisons between treatment conditions.

Pairwise comparisons between treatments indicate clear egocentric effects. In line with H1, creators evaluated their own creations more highly than invested and uninvested evaluators judging the same images (\diff{1.743} and \diff{1.986}, respectively). As hypothesized by H2, invested evaluators judged their own images more positively than those created by other participants (\diff{1.565}). We found that uninvested evaluators score images generated by creators lower than invested evaluators score their own creations (\diff{1.808}), as suggested by H4. {Finally, we also observed significant order effects. The more images a participant evaluated, the lower they scored images ($b$ = -0.094, $SE$ = 0.046, \ttest{1533}{-2.036}{0.05}).}

\section{Discussion}
\label{sec:discussion}
\subsection{Laypeople’s Perceptions of AI-Generated Art in Relation to Copyright}

\subsubsection{RQ1: Creativity and Effort---Not Skills---Are Required To Generate Art with AI}

Participants believed that creativity and effort were necessary to generate images with the GenAI model. In contrast, they neither agreed nor disagreed that creators used their skills when creating AI art. {In other words, when we analyze how participants perceived the nature of the digital artifacts produced by GenAI through the lens of these copyright-relevant factors, we find that people believe AI-generated art reproduces its creator's creativity and effort, even though producing it may not require advanced artistic skills.} 

The fact that the AI generation process was described as creative seems to suggest that AI art could have the modicum of creativity that US copyright law requires for a work to warrant copyright~\cite{us2023copyright}, which is at odds with current decisions to reject protection for AI-generated images~\cite{copyrightverge}. Concerning judgments of effort, we note that our study provided evaluators with information that can work as a proxy for effort (e.g., the total number of images creators generated and how long they spent in the study), which could have influenced how participants judged the effort put by creators. Nonetheless, our experimental findings are robust to the inclusion of these covariates. That is, our findings remain qualitatively the same even when accounting for the aforementioned proxies for effort. Finally, perceptions of skills could have been limited by how much control creators had over the images they generated. Had creators been allowed to edit the GenAI outputs, evaluators could have judged the creation process as requiring more skills---a research direction that future work could explore.

\subsubsection{RQ2: Users and Data Contributors as Authors}

Users and data contributors were attributed the most authorship for AI-generated art. Although it is less surprising that those who prompted the GenAI model to generate images are perceived as authors, participants' acknowledgment of data contributors is noteworthy. In line with prior work suggesting that laypeople agree that data contributors should be compensated if their work is used for training GenAI models~\cite{vergesurvey}, our results suggest that current practices that fail to compensate and appreciate data contributors---which some equate to theft~\cite{stealingnewyorker,merkley}---may be at odds with laypeople's expectations. People's opinions seem to be more aligned with proposals of licensing models, under which data contributors can be compensated for the use of their creations~\cite{licensing}.

Our study explained to participants that GenAI models learn from a ``dataset of human-created content,'' which could have made the role of data contributors more prominent in our study, potentially influencing how people perceived their role in GenAI. Nonetheless, it is important to emphasize that GenAI models require such datasets to work, meaning that any explanation of how GenAI works without mentioning training data is incomplete. We call for future work exploring how different ways of introducing data contributors impact how much authorship and credit laypeople grant to those whose creations are used to train AI.

{By designing systems that fail to compensate those whose creations are needed to train GenAI, system designers may end up perpetuating existing patterns of---what some authors would call---exploitation~\cite{stealingnewyorker,merkley}. Worryingly, past work suggests that even when developers want to question the implications of their products, they often lack the power to contest organizational dynamics and incentives~\cite{widder2023s,widder2024power}. Future work could bring together prior research developing mechanisms for collective action~\cite{krafft2021action,calacci2022bargaining} together with self-questioning design practice and education on the potential ethical concerns posed by GenAI~\cite{garrett2020more} to identify mechanisms through which HCI practitioners and researchers can resist entrenchment of exploitive design practices.}

People's attribution of authorship to training data contributors also highlights the importance of CS research for copyright law in the age of GenAI. To adequately compensate training data contributors for the impact their work had on a generated image, it is important to be able to identify which data points contributed to the generation of the image, as well as estimate their own individual contribution. This is closely related to the problem of data valuation in the machine learning literature~\cite{ghorbani2019data, jia2019towards,pruthi2020estimating,park2023trak}. These estimations, however, have been found to be intractable in certain circumstances~\cite{hammoudeh2024training}. Hence, we call for further research on how to robustly quantify training data influence, which can provide technical solutions for compensating training data contributors appropriately.

Our result that users were acknowledged as authors of the images they generate using AI calls for the reconsideration of legal decisions that have refused to grant copyright to GenAI outputs in the US~\cite{copyrightverge}. The main reason given for the rejection was that AI-generated images do not have human authors, making them ineligible for copyright protection. In contrast, our findings suggest that laypeople believe users are authors, even in a scenario in which they were not allowed to edit the AI-generated images. We expect the perceived authorship of users to increase when given the chance to exert more control over GenAI outputs.

Participants were uncertain about the authorship of the AI model, which is aligned with prior work suggesting that laypeople are neither against nor in favor of copyright rights to AI models~\cite{lima2020collecting}. Finally, respondents' disagreement with the idea of companies being authors of GenAI outputs is at odds with potential interpretations of UK IP law, which could be interpreted as the companies training GenAI models holding copyright over outputs~\cite{ihalainen2018computer}. That is, current practices that concentrate profits in the hands of the corporations training GenAI models without little to no compensation to data contributors~\cite{stealingnewyorker,merkley} do not align with lay opinions about copyright of AI-generated art. 

We note, however, that these two actors, to whom participants attributed lower levels of authorship, are non-human entities. In contrast, the user and data contributors are human, suggesting that our findings could have been influenced by the human nature of these actors. Future work could replicate our study by also investigating whether the (human) programmers who developed the model would be granted authorship, which could indicate that the human nature of potential authors plays a role in people's perceptions of authorship.

\subsubsection{RQ3: Users and Data Contributors as Rights-Holders}

Similar to our results concerning authorship, those who prompted the AI model to generate images (i.e., users) and data contributors were frequently identified as potential rights-holders of AI-generated art. {All in all, we find that people grant ownership of AI-generated digital artifacts to users and those who contributed creatively to the system design and not necessarily to the technological artifact itself and the entity who developed it.}

Our findings have two main legal implications. First, it puts into question the exclusive nature of the rights afforded by copyright protection; participants seemed to support a more distributed approach, under which not only ``authors'' would be owners of the copyright, but also some actors without which AI-generated art would not be possible. For instance, a joint ownership model~\cite{khosrowi2023diffusing} could be more aligned with laypeople's expectations regarding AI-generated art. Second, it highlights participants' calls for the compensation of data contributors. An approach that could be more aligned with lay opinions could rely on neighboring rights~\cite{lee2021computer,iaia2022or}, which would still center ownership around ``authors'' (i.e., users) but without neglecting the interests of data contributors, e.g., through the distribution of royalties. As discussed above, another solution could be based on licensing models for training data~\cite{licensing}.

Participants were more supportive of non-exclusive rights over AI-generated art in non-commercial settings, which is in line with doctrines that permit the use of copyrighted material for non-commercial purposes (e.g., the US fair use doctrine). Future debates concerning the use of AI-generated outputs in relation to copyright could consider our results to determine whether their use should be legal.

{Although copyright law worldwide does not yet have a definitive answer to whether AI-generated art is copyrightable, another legally relevant document that might determine who might own outputs is each GenAI model's terms of service. Even though most terms of service state that users own the content they create ``to the extent permitted by applicable law''~\cite{chatgptterms,midjourneyterms,stabilityterms}, they also grant companies the right to use the AI-generated content to improve their models. Such terms could acknowledge the role of data contributors, restrict the rights of developing companies, and consider whether the content is being used commercially or not.\footnote{{Midjourney~\cite{midjourneyterms}, for instance, only grants ownership of AI outputs to employees of companies with more than \$1,000,000 USD if they are subscribed to one of their premium plans.}}}

\subsection{Egocentric Biases (Or Lack Thereof) in Perceptions of AI-Generated Images}

\subsubsection{Creativity, Effort, and Skills}

Judgments regarding creativity and effort partially exhibited egocentric effects. Creators attributed more creativity to their creations than uninvested evaluators (partial H1), in line with prior work~\cite{buccafusco2011creativity}. Evaluations of the effort put into generating images with GenAI are also consistent with our egocentric hypotheses (partial H1 and H2). Furthermore, judgments concerning creativity are aligned with H4 (i.e., experience effect), which suggests that people may judge images more favorably after interacting with the GenAI model. Judgments about effort and skills are also partially aligned with H4.

Surprisingly, our results concerning perceived creativity and skills suggest the opposite of the hypothesized competition effect (H3). Even though invested evaluators were judging images against which they were competing in the exhibition, they assigned higher creativity and skill ratings than uninvested evaluators (i.e., those not competing for the monetary award). 

With the current experimental design, it is difficult to distinguish between egocentric and experience effects since all participants who had reasons to judge images egocentrically also interacted with the model. Future work could disentangle these two effects. For instance, some participants could enroll in the competition with an image that they did not generate (e.g., with images generated by the researchers). This treatment condition would maintain potential egocentric effects, since participants would still be incentivized to favor their own submissions to the exhibition, without interacting with the GenAI model, mitigating experience effects. 

Another approach that could help research understand how people perceive AI-generated art in relation to copyright is exploring whether the ``experience effect'' (H4) emerges from interacting with the model or competing in our AI art exhibition. In our study, all participants who interacted with the model also submitted an image for consideration at the exhibition. Future work could distinguish between these two factors by having a condition in which creators use the GenAI model to generate images without the competition setting, thus capturing only the effect of using the model on people's perceptions.

\subsubsection{Perceived Authorship}

We hypothesized that participants would exhibit egocentric biases in their attribution of authorship to users. However, our findings are at odds with our hypotheses. On average, invested evaluators attributed more authorship to other people compared to themselves, which is contrary to H2. When evaluating other people's images, invested evaluators also credited users as authors more than the users themselves (i.e., creators), a result that is also misaligned with H1. 

A potential explanation for why creators attributed less authorship to themselves than to other people is that authorship does not only encode positive rights but also negative responsibilities from which creators tried to distance themselves. Although authorship is often associated with positive outcomes in the context of copyright law (e.g., the opportunity to profit from one's creation), it could also lead to the responsibility for any negative outcomes or backlash arising from their work being published~\cite{nehamas1986author,foucault2019aesthetics}. If one is an author under this interpretation, they are also responsible, for instance, if their creation is found to be infringing on another copyrighted work, which is relevant for debates concerning AI-generated images. Hence, creators seem to want to enjoy the rights associated with copyright (see above) but might rather not be perceived as authors because of potential negative repercussions in case something goes wrong. Future work could explore this potential explanation further. As suggested by prior work showing that users can escape responsibility by delegating tasks to AI systems~\cite{feier2022hiding,kobis2021bad}, studies could examine how creators distance themselves from AI-generated art when negative responsibilities are salient.

{Participants' distancing from potential negative outcomes of using GenAI is in conflict with current terms of service. Existing terms state that users are fully responsible for whatever they generate using the model, including if outputs infringe on the copyright of any third party~\cite{chatgptterms,stabilityterms,midjourneyterms}. Companies developing GenAI models could consider decreasing (or sharing) the potential liability of users in their terms of service to align with users' expectations.}

In H4, we hypothesized that having experience with the AI model would influence participants' attribution of authorship to the user. Informally, we expected that by interacting with the AI model, people would notice the degree of influence a user's prompts have over the model's outputs, leading them to attribute more authorship to users. We did not, however, find a significant difference between uninvested evaluators and other groups of participants in terms of perceptions of user authorship. It is important to note that H4 focused on the experience of observing the relationship between one's own inputs and the model's outputs. Perhaps there was not much variance in each participant's inputs, or they generated too few images (the average participant prompted the model a median of two times) for the experience effects to kick in.

However, we identified a different effect of experience. The more images a participant evaluated, the more authorship they attributed to users, suggesting order effects. The identified order effects relate to a different form of experience: observing the relationship between other participants' inputs and model outputs. It is likely that there was more between-participants variance than within-participant variations in the images they generated.\footnote{This explanation would be aligned with prior research in social psychology, which finds higher degrees of variance in the judgments of different people than between the judgments of the same person in different points in time~\cite{kahneman2016noise, kahneman2021noise}.} Since evaluators observed the prompts from four different creators, they had more opportunities for this effect of experience to come into play.

A potential explanation connecting our findings regarding H1, H2, H4, and order effects is that being exposed to diverse GenAI inputs and outputs makes evaluators value the human component of AI-generated images more. After using the GenAI model, seeing what other participants generated, and realizing that different human actions led to widely different outputs, invested evaluators may have started acknowledging that AI cannot generate images without some level of human input and authorship. Creators and invested evaluators judged their own images without such experience, while invested evaluators judged the creators' images after this experience, hence possibly explaining why the latter received higher authorship scores (opposite of H1 and H2). To explore this potential explanation, future work could explore inverting the order in which invested evaluators generate and evaluate images. By first evaluating other people's submissions and then generating their own image, participants might end up overestimating their authorship after seeing what others were able to create, providing further evidence of order effects.

Finally, we note that the increase in the perceived authorship of the user was accompanied by a slight decrease in AI model's authorship. It is possible that seeing how different participants generated images of varying quality---while using the \emph{same} GenAI model---highlighted the skills behind prompting, leading to higher perceived authorship to users and slightly lower perceived authorship of the AI model.

\subsubsection{Attribution of Rights}

Finally, we also identified some differences between treatments regarding the right to display AI-generated images: creators were more likely to support non-exclusive rights for their own images than invested evaluators evaluating the same images. A potential explanation is that creators would like audiences to see their creations because it could lead to further financial benefits in the long run. Instead of keeping their creations ``behind curtains,'' creators could have imagined that allowing others to display their images could bring financial benefits down the road. Additional longitudinal studies are necessary to explore this hypothesis further.

\subsubsection{Actual Rewards vs. Hypothetical Rights}

Participants evaluated their own creations egocentrically when this assessment was supposed to determine who received monetary awards. On average, creators rated their images approximately 1.9 points higher (on an 11-point scale) than evaluators evaluating the same images. Furthermore, invested evaluators evaluated their own images around 1.5 points higher than they evaluated creators' images. Our results suggest that financial incentives can bias people's responses to AI-generated art. However, when asked who should hypothetically have some of the rights associated with copyright protection, participants did not prioritize their own gains. Instead, egocentric biases emerged only in the context of actual rewards, and not hypothetical rights, suggesting that future debates concerning AI-generated works in relation to copyright law may be tainted by conflicts of interest.

Participants' tendency to rate their creations egocentrically when financial incentives are present but not when asked about hypothetical rights is also relevant in relation to data contributors. We found that people acknowledged data contributors as authors and potential rights-holders; yet, it is unclear whether we would observe the same result had the corresponding questions involved actual monetary awards. It is possible that if participants were asked to \emph{financially} compensate data contributors, e.g., by sharing their award, people would not do it and would rather keep the incentive to themselves. Such a finding would suggest that laypeople may support regulatory frameworks that compensate data contributors only if these frameworks do not affect them financially. Our study design could be easily modified to address these research questions. For instance, future studies could include questions concerning participants' willingness to share their potential reward with data contributors.

It is worth noting that while creators and invested evaluators assigned higher ratings to their own images, we found no evidence of the sabotage hypothesized in H3. That is, invested evaluators did not give creators significantly lower scores than uninvested evaluators. It is unclear whether this lack of sabotage is the result of a lack of understanding of the incentive structure or if it reflects participants' true altruistic behavior. Future research is needed to test participants' understanding of the incentive structure. Another potential explanation is that participants chose to inflate their own scores instead of lowering other participants' scores because doing so could be seen as less morally contentious---a hypothesis that future work could explore further.

Our finding that granting monetary awards has the potential to make people judge AI-generated images egocentrically has implications for copyright law. Copyright law relies on similar financial incentives to fulfill its normative goal of promoting creativity. Extending copyright to a work of authorship restricts who can benefit and profit from it---i.e., decides who receives the financial incentives associated with copyright---thus making potential rights-holders susceptible to egocentric biases.

{The fact that the potential of monetary awards changed people's perceptions of GenAI outputs also has implications for the design of AI models. There are already examples of how AI-generated art could give users financial benefits~\cite{nytexhibition} even though its legal status is still undefined. Future work could explore design patterns for GenAI products that are more effectively transparent about potential legal and financial consequences so that users are aware of both positive and negative outcomes (as discussed above in relation to perceived authorship).}

\subsection{Concluding Remarks}

In summary, we found that users of GenAI judge their own creations egocentrically concerning some factors but not others. In fact, our results even suggest effects in the opposite direction for some of the variables we examined (e.g., perceived authorship of the user). Importantly, we identified the importance of financial interests in determining whether people favor their own AI-generated images compared to other people's creations.

This research focused on how laypeople perceive AI-generated images in relation to copyright. However, copyright law regulates other works of authorship that could challenge the law if generated with GenAI. For instance, the legal status of AI-generated songs and novels has also fueled contentious debates~\cite{AIsong,AInovel}. We explored images because of the wide availability of AI image models, as well as its potential lower cognitive load for participants compared to other works of authorship, such as novels, which would take longer to evaluate. We call for future work investigating how laypeople's expectations of copyright for AI-generated works may vary depending on the content form.

All the factors we explore in this study impact whether something is eligible for copyright protection. In other words, a comprehensive analysis of all these factors is what determines whether a work of authorship, AI-generated or otherwise, is protected under copyright law. Our findings that some of these factors are susceptible to egocentric biases---while others are not---raise the question of whether current methods of determining copyright eligibility are appropriate for GenAI. Future work could explore whether some of these factors are indeed appropriate in the context of GenAI, as well as examine whether other variables that are currently irrelevant in determinations regarding human-created works should be considered when examining AI-generated outputs.

\begin{acks}

This work was supported by the Max Planck Institute for Software Systems; the first and third author have completed a portion of this work while at the Max Planck Institute for Software Systems. We are grateful for the feedback provided by Toni-Lee Sangastiano, Kristelia García, members of the Berkman Klein Center for Internet \& Society at Harvard University and the working groups on Generative AI and Copyright and AI Governance held there, and the attendees of the workshop on Experimental Methods in Legal Scholarship held at the Max Planck Institute for Research on Collective Goods.

\end{acks}

\bibliographystyle{ACM-Reference-Format}
\bibliography{sample-base}

\newpage

\appendix

\section{Additional Background on Copyright and Generative AI}

{Below, we provide additional background on some of the debates surrounding GenAI and copyright law. More specifically, we introduce the literature exploring whether training GenAI models violates the copyright of the training data and discuss some potential alternative regulatory frameworks for GenAI. See Section~\ref{sec:background} for debates at the intersection of GenAI and copyright law that are directly relevant to the study.}

\subsection{Does Training GenAI Infringe on the Copyright of the Training Data?}

{GenAI models require large amounts of data to be trained. These datasets may contain copyrighted data, raising the question of whether the training of AI models infringes on the rights of copyright holders. Those arguing that training GenAI on copyrighted material should be illegal often posit that it exploits authors of the training data without compensating them~\cite{choi2023protecting}, with some going even further and equating the practice to theft~\cite{stealingnewyorker,merkley}. These critics often defend that the owners of training data should be compensated~\cite{chan2023reclaiming}.}

{In contrast, proponents of GenAI assert that training does not infringe on the copyright of its training data. Such arguments in favor of GenAI often rely on the United States' fair use doctrine~\cite{lemley2020fair}, which permits limited use of copyrighted materials under specific conditions, and the text and data mining copyright exceptions in the European Union (EU)~\cite{selvadurai2020reconsidering}. The wide availability of GenAI models has triggered several lawsuits that are currently underway (e.g., ~\cite{nytlawsuit}) and will decide whether training GenAI with copyrighted material without compensating its owners is legal under copyright law.}

\subsection{Alternatives to Copyright Law}

{Although copyright law often takes central stage in the discussion surrounding ownership of GenAI outputs, it is not exempt from critiques. Copyright law is often criticized for its potential to hinder innovation by concentrating power in monopolies~\cite{ihalainen2018computer,levendowski2018copyright}. Furthermore, extending copyright law to GenAI could conflict with its primary objective---to promote \emph{human} creativity; if AI-generated works are protected, they could devalue human creativity by flooding the market with artificial competition to human creations~\cite{iaia2022or}.}

{Scholars have proposed different approaches for AI-generated works to promote innovation. For instance, the reproduction of GenAI models and their outputs could be restricted under a \textit{suis generis} doctrine, which grants exclusive rights in situations in which there has been substantial financial investment~\cite{smits2022generative}. Someone other than the author---which is hard to define in the case of AI-generative works---could also have ``neighbouring'' rights over AI outputs as a way to protect their investment~\cite{lee2021computer,iaia2022or}. AI-generated outputs could also be covered by other branches of IP law, such as trademark law, which relies on owners maintaining and enforcing their rights~\cite{ihalainen2018computer}. Another approach could be moving away from copyright law's focus on creativity and originality as grounds for ownership, and instead reward actors who put effort and demonstrate skill when using GenAI~\cite{guadamuz2017androids}.}

\section{Supplementary Analysis}

{We present robustness tests of our regression results, controlling for order effects, participant-level, and image-level variables in Tables~\ref{tab:robustness_creativity}-\ref{tab:robustness_exhibition} (see Section~\ref{sec:analysis_plan} for details). Our findings are robust to the inclusion of these covariates, with only minor changes to the significance of borderline contrasts. Tables~\ref{tab:dp_pairwise}-\ref{tab:cn_pairwise} present the pairwise comparisons of attribution of rights between treatment conditions. We omit these tables from the main text for conciseness since we did not find support for any of our hypotheses regarding attribution of rights to users.}


\begin{table*}[!htbp] \centering 
\small
\begin{tabular}{@{\extracolsep{5pt}}lcccc} 
\\[-1.8ex]\hline 
\hline \\[-1.8ex] 
 & \multicolumn{4}{c}{\textit{Creativity}} \\ 
\cline{2-5} 
\\[-1.8ex] & (1) & (2) & (3) & (4)\\ 
\hline \\[-1.8ex] 
 Invested Evaluators (Own) & $-$0.140 & $-$0.140 & $-$0.277 & $-$0.144 \\ 
  & (0.206) & (0.206) & (0.210) & (0.197) \\ 

 Invested Evaluators (Others) & $-$0.434$^{*}$ & $-$0.428$^{*}$ & $-$0.567$^{**}$ & $-$0.410$^{*}$ \\ 
  & (0.178) & (0.183) & (0.186) & (0.181) \\ 

 Uninvested Evaluators (Others)  & $-$0.900$^{***}$ & $-$0.897$^{***}$ & $-$0.933$^{***}$ & $-$0.924$^{***}$ \\ 
  & (0.191) & (0.190) & (0.194) & (0.189) \\ 

 Evaluation order &  & $-$0.002 &  &  \\ 
  &  & (0.029) &  &  \\ 

 Frequency of using GenAI for images &  &  & 0.313$^{***}$ &  \\ 
  &  &  & (0.083) &  \\ 

 Frequency of using GenAI for text &  &  & 0.048 &  \\ 
  &  &  & (0.077) &  \\ 

 Previous participation in GenAI generation study &  &  & 0.170 &  \\ 
  &  &  & (0.156) &  \\ 

 Previous participation in GenAI evaluation study &  &  & $-$0.107 &  \\ 
  &  &  & (0.173) &  \\ 

 Art training &  &  & $-$0.103 &  \\ 
  &  &  & (0.159) &  \\ 

 Computer science training &  &  & 0.008 &  \\ 
  &  &  & (0.165) &  \\ 

 Law training &  &  & $-$0.140 &  \\ 
  &  &  & (0.220) &  \\ 

 Age &  &  & 0.016$^{**}$ &  \\ 
  &  &  & (0.005) &  \\ 

 Gender = Non-binary/Prefer not to respond  &  &  & 0.063 &  \\ 
  &  &  & (0.354) &  \\ 

 Gender = Woman &  &  & $-$0.161 &  \\ 
  &  &  & (0.133) &  \\ 

 Race = Black/African American &  &  & $-$0.180 &  \\ 
  &  &  & (0.240) &  \\ 

 Race = Other/Mixed/Prefer not to respond &  &  & 0.166 &  \\ 
  &  &  & (0.234) &  \\ 

 Race = White &  &  & $-$0.023 &  \\ 
  &  &  & (0.195) &  \\ 

 Number of images images generated before final selection &  &  &  & 0.015 \\ 
  &  &  &  & (0.012) \\ 

 Time until selecting image &  &  &  & 0.0004$^{*}$ \\ 
  &  &  &  & (0.0002) \\ 

 Length of prompt (in characters) &  &  &  & 0.004$^{***}$ \\ 
  &  &  &  & (0.001) \\ 

 Image type = Landscape &  &  &  & 0.198 \\ 
  &  &  &  & (0.150) \\ 

 image type = Portrait &  &  &  & 0.359$^{**}$ \\ 
  &  &  &  & (0.138) \\ 

 Constant & 1.211$^{***}$ & 1.211$^{***}$ & 0.373 & 0.418 \\ 
  & (0.163) & (0.163) & (0.319) & (0.226) \\ 
  & & & & \\

\hline
\hline 
\end{tabular} 
  \caption{\textbf{Robustness tests of our results on perceived creativity.} We include three sets of covariates and find consistent experimental effects. Model (1) is the baseline model without any covariates; model (2) includes a variable indicating the order in which the image was shown to an evaluator to identity potential order effects; model (3) includes participant-level covariates; and model (4) includes image-level covariates. Refer to Section~\ref{sec:analysis_plan} for details. Standard errors are shown inside parentheses. $^{*}$p$<$0.05; $^{**}$p$<$0.01; $^{***}$p$<$0.001.} 
  \label{tab:robustness_creativity} 
\end{table*}

\begin{table*}[!htbp] \centering 
\small
\begin{tabular}{@{\extracolsep{5pt}}lcccc} 
\\[-1.8ex]\hline 
\hline \\[-1.8ex] 
 & \multicolumn{4}{c}{\textit{Effort}} \\ 
\cline{2-5} 
\\[-1.8ex] & (1) & (2) & (3) & (4)\\ 
\hline \\[-1.8ex] 
  Invested Evaluators (Own) & 0.084 & 0.084 & $-$0.092 & 0.102 \\ 
  & (0.227) & (0.227) & (0.233) & (0.214) \\ 

 Invested Evaluators (Others) & $-$0.373 & $-$0.407 & $-$0.545$^{*}$ & $-$0.342 \\ 
  & (0.206) & (0.223) & (0.212) & (0.209) \\ 

 Uninvested Evaluators (Others)  & $-$0.754$^{***}$ & $-$0.775$^{***}$ & $-$0.814$^{***}$ & $-$0.780$^{***}$ \\ 
  & (0.193) & (0.197) & (0.194) & (0.193) \\ 

 Evaluation order &  & 0.014 &  &  \\ 
  &  & (0.035) &  &  \\ 

 Frequency of using GenAI for images &  &  & 0.274$^{**}$ &  \\ 
  &  &  & (0.087) &  \\ 

 Frequency of using GenAI for text &  &  & 0.176$^{*}$ &  \\ 
  &  &  & (0.083) &  \\ 

 Previous participation in GenAI generation study &  &  & 0.227 &  \\ 
  &  &  & (0.167) &  \\ 

 Previous participation in GenAI evaluation study &  &  & $-$0.110 &  \\ 
  &  &  & (0.172) &  \\ 

 Art training &  &  & $-$0.168 &  \\ 
  &  &  & (0.164) &  \\ 

 Computer science training &  &  & 0.105 &  \\ 
  &  &  & (0.162) &  \\ 

 Law training &  &  & $-$0.199 &  \\ 
  &  &  & (0.237) &  \\ 

 Age &  &  & 0.022$^{***}$ &  \\ 
  &  &  & (0.005) &  \\ 

 Gender = Non-binary/Prefer not to respond  &  &  & 0.242 &  \\ 
  &  &  & (0.360) &  \\ 

 Gender = Woman &  &  & $-$0.095 &  \\ 
  &  &  & (0.142) &  \\ 

 Race = Black/African American &  &  & $-$0.269 &  \\ 
  &  &  & (0.239) &  \\ 

 Race = Other/Mixed/Prefer not to respond &  &  & 0.139 &  \\ 
  &  &  & (0.241) &  \\ 

 Race = White &  &  & $-$0.059 &  \\ 
  &  &  & (0.185) &  \\ 

 Number of images images generated before final selection &  &  &  & 0.033$^{**}$ \\ 
  &  &  &  & (0.012) \\ 

 Time until selecting image &  &  &  & 0.0004$^{*}$ \\ 
  &  &  &  & (0.0002) \\ 

 Length of prompt (in characters) &  &  &  & 0.004$^{***}$ \\ 
  &  &  &  & (0.001) \\ 

 Image type = Landscape &  &  &  & 0.228 \\ 
  &  &  &  & (0.135) \\ 

 image type = Portrait &  &  &  & 0.267 \\ 
  &  &  &  & (0.139) \\ 

 Constant & 0.916$^{***}$ & 0.916$^{***}$ & $-$0.302 & 0.018 \\ 
  & (0.182) & (0.182) & (0.341) & (0.200) \\

\hline 
\hline 
\end{tabular} 
  \caption{\textbf{Robustness tests of our results on perceived effort.} We include three sets of covariates and find consistent experimental effects. Model (1) is the baseline model without any covariates; model (2) includes a variable indicating the order in which the image was shown to an evaluator to identity potential order effects; model (3) includes participant-level covariates; and model (4) includes image-level covariates. Refer to Section~\ref{sec:analysis_plan} for details. Standard errors are shown inside parentheses. $^{*}$p$<$0.05; $^{**}$p$<$0.01; $^{***}$p$<$0.001.} 
  \label{tab:robustness_effort} 
\end{table*}

\begin{table*}[!htbp] \centering 
\small
\begin{tabular}{@{\extracolsep{5pt}}lcccc} 
\\[-1.8ex]\hline 
\hline \\[-1.8ex] 
 & \multicolumn{4}{c}{\textit{Skill}} \\ 
\cline{2-5} 
\\[-1.8ex] & (1) & (2) & (3) & (4)\\ 
\hline \\[-1.8ex] 
 Invested Evaluators (Own) & 0.059 & 0.059 & $-$0.202 & 0.045 \\ 
  & (0.236) & (0.236) & (0.238) & (0.231) \\ 

 Invested Evaluators (Others) & 0.249 & 0.207 & $-$0.006 & 0.274 \\ 
  & (0.227) & (0.240) & (0.231) & (0.229) \\ 

 Uninvested Evaluators (Others)  & $-$0.245 & $-$0.270 & $-$0.339 & $-$0.266 \\ 
  & (0.212) & (0.214) & (0.211) & (0.212) \\ 

 Evaluation order &  & 0.017 &  &  \\ 
  &  & (0.034) &  &  \\ 

 Frequency of using GenAI for images &  &  & 0.368$^{***}$ &  \\ 
  &  &  & (0.094) &  \\ 

 Frequency of using GenAI for text &  &  & 0.233$^{**}$ &  \\ 
  &  &  & (0.085) &  \\ 

 Previous participation in GenAI generation study &  &  & 0.180 &  \\ 
  &  &  & (0.201) &  \\ 

 Previous participation in GenAI evaluation study &  &  & $-$0.094 &  \\ 
  &  &  & (0.201) &  \\ 

 Art training &  &  & $-$0.193 &  \\ 
  &  &  & (0.186) &  \\ 

 Computer science training &  &  & 0.031 &  \\ 
  &  &  & (0.186) &  \\ 

 Law training &  &  & 0.069 &  \\ 
  &  &  & (0.260) &  \\ 

 Age &  &  & 0.026$^{***}$ &  \\ 
  &  &  & (0.006) &  \\ 

 Gender = Non-binary/Prefer not to respond  &  &  & $-$0.144 &  \\ 
  &  &  & (0.421) &  \\ 

 Gender = Woman &  &  & $-$0.266 &  \\ 
  &  &  & (0.150) &  \\ 

 Race = Black/African American &  &  & $-$0.081 &  \\ 
  &  &  & (0.265) &  \\ 

 Race = Other/Mixed/Prefer not to respond &  &  & 0.210 &  \\ 
  &  &  & (0.290) &  \\ 

 Race = White &  &  & $-$0.055 &  \\ 
  &  &  & (0.221) &  \\ 

 Number of images images generated before final selection &  &  &  & 0.007 \\ 
  &  &  &  & (0.012) \\ 

 Time until selecting image &  &  &  & 0.001$^{**}$ \\ 
  &  &  &  & (0.0002) \\ 

 Length of prompt (in characters) &  &  &  & 0.003$^{***}$ \\ 
  &  &  &  & (0.001) \\ 

 Image type = Landscape &  &  &  & 0.129 \\ 
  &  &  &  & (0.122) \\ 

 image type = Portrait &  &  &  & 0.156 \\ 
  &  &  &  & (0.128) \\ 

 Constant & $-$0.095 & $-$0.095 & $-$1.566$^{***}$ & $-$0.742$^{***}$ \\ 
  & (0.184) & (0.184) & (0.379) & (0.198) \\

\hline
\hline 
\end{tabular} 
  \caption{\textbf{Robustness tests of our results on perceived skill.} We include three sets of covariates and find consistent experimental effects. Model (1) is the baseline model without any covariates; model (2) includes a variable indicating the order in which the image was shown to an evaluator to identity potential order effects; model (3) includes participant-level covariates; and model (4) includes image-level covariates. Refer to Section~\ref{sec:analysis_plan} for details. Standard errors are shown inside parentheses. $^{*}$p$<$0.05; $^{**}$p$<$0.01; $^{***}$p$<$0.001.} 
  \label{tab:robustness_skill} 
\end{table*}

\begin{table*}[!htbp] \centering 
\small
\begin{tabular}{@{\extracolsep{5pt}}lcccc} 
\\[-1.8ex]\hline 
\hline \\[-1.8ex] 
 & \multicolumn{4}{c}{\textit{User Authorship}} \\ 
\cline{2-5} 
\\[-1.8ex] & (1) & (2) & (3) & (4)\\ 
\hline \\[-1.8ex] 
  Invested Evaluators (Own) & 0.199 & 0.199 & $-$0.047 & 0.175 \\ 
  & (0.218) & (0.218) & (0.228) & (0.216) \\ 

 Invested Evaluators (Others) & 0.845$^{***}$ & 0.742$^{***}$ & 0.605$^{**}$ & 0.857$^{***}$ \\ 
  & (0.210) & (0.212) & (0.218) & (0.210) \\ 

 Uninvested Evaluators (Others)  & 0.382 & 0.320 & 0.282 & 0.374 \\ 
  & (0.206) & (0.206) & (0.206) & (0.206) \\ 

 Evaluation order &  & 0.041 &  &  \\ 
  &  & (0.022) &  &  \\ 

 Frequency of using GenAI for images &  &  & 0.187 &  \\ 
  &  &  & (0.102) &  \\ 

 Frequency of using GenAI for text &  &  & 0.215$^{*}$ &  \\ 
  &  &  & (0.097) &  \\ 

 Previous participation in GenAI generation study &  &  & 0.061 &  \\ 
  &  &  & (0.214) &  \\ 

 Previous participation in GenAI evaluation study &  &  & $-$0.218 &  \\ 
  &  &  & (0.222) &  \\ 

 Art training &  &  & 0.108 &  \\ 
  &  &  & (0.180) &  \\ 

 Computer science training &  &  & $-$0.014 &  \\ 
  &  &  & (0.186) &  \\ 

 Law training &  &  & 0.101 &  \\ 
  &  &  & (0.244) &  \\ 

 Age &  &  & 0.026$^{***}$ &  \\ 
  &  &  & (0.006) &  \\ 

 Gender = Non-binary/Prefer not to respond  &  &  & 0.036 &  \\ 
  &  &  & (0.434) &  \\ 

 Gender = Woman &  &  & $-$0.144 &  \\ 
  &  &  & (0.165) &  \\ 

 Race = Black/African American &  &  & 0.174 &  \\ 
  &  &  & (0.239) &  \\ 

 Race = Other/Mixed/Prefer not to respond &  &  & 0.255 &  \\ 
  &  &  & (0.271) &  \\ 

 Race = White &  &  & $-$0.043 &  \\ 
  &  &  & (0.198) &  \\ 

 Number of images images generated before final selection &  &  &  & $-$0.001 \\ 
  &  &  &  & (0.009) \\ 

 Time until selecting image &  &  &  & 0.0003 \\ 
  &  &  &  & (0.0001) \\ 

 Length of prompt (in characters) &  &  &  & 0.001$^{**}$ \\ 
  &  &  &  & (0.001) \\ 

 Image type = Landscape &  &  &  & $-$0.048 \\ 
  &  &  &  & (0.110) \\ 

 image type = Portrait &  &  &  & $-$0.041 \\ 
  &  &  &  & (0.123) \\ 

 Constant & $-$0.021 & $-$0.021 & $-$1.405$^{***}$ & $-$0.217 \\ 
  & (0.172) & (0.172) & (0.343) & (0.198) \\

\hline
\hline 
\end{tabular} 
  \caption{\textbf{Robustness tests of our results on user authorship.} We include three sets of covariates and find consistent experimental effects. Model (1) is the baseline model without any covariates; model (2) includes a variable indicating the order in which the image was shown to an evaluator to identity potential order effects; model (3) includes participant-level covariates; and model (4) includes image-level covariates. Refer to Section~\ref{sec:analysis_plan} for details. Standard errors are shown inside parentheses. $^{*}$p$<$0.05; $^{**}$p$<$0.01; $^{***}$p$<$0.001.} 
  \label{tab:robustness_user} 
\end{table*} 

\begin{table*}[!htbp] \centering 
\small
\begin{tabular}{@{\extracolsep{5pt}}lcccc} 
\\[-1.8ex]\hline 
\hline \\[-1.8ex] 
 & \multicolumn{4}{c}{\textit{AI Model Authorship}} \\ 
\cline{2-5} 
\\[-1.8ex] & (1) & (2) & (3) & (4)\\ 
\hline \\[-1.8ex] 
  Invested Evaluators (Own) & $-$0.295 & $-$0.295 & $-$0.328 & $-$0.292 \\ 
  & (0.225) & (0.225) & (0.237) & (0.223) \\ 

 Invested Evaluators (Others) & $-$0.470$^{*}$ & $-$0.311 & $-$0.502$^{*}$ & $-$0.469$^{*}$ \\ 
  & (0.220) & (0.229) & (0.229) & (0.221) \\ 

 Uninvested Evaluators (Others)  & $-$0.319 & $-$0.224 & $-$0.364 & $-$0.315 \\ 
  & (0.221) & (0.224) & (0.223) & (0.221) \\ 

 Evaluation order &  & $-$0.064$^{*}$ &  &  \\ 
  &  & (0.028) &  &  \\ 

 Frequency of using GenAI for images &  &  & 0.033 &  \\ 
  &  &  & (0.110) &  \\ 

 Frequency of using GenAI for text &  &  & 0.349$^{***}$ &  \\ 
  &  &  & (0.104) &  \\ 

 Previous participation in GenAI generation study &  &  & $-$0.080 &  \\ 
  &  &  & (0.246) &  \\ 

 Previous participation in GenAI evaluation study &  &  & 0.118 &  \\ 
  &  &  & (0.244) &  \\ 

 Art training &  &  & $-$0.338 &  \\ 
  &  &  & (0.200) &  \\ 

 Computer science training &  &  & 0.151 &  \\ 
  &  &  & (0.217) &  \\ 

 Law training &  &  & 0.191 &  \\ 
  &  &  & (0.330) &  \\ 

 Age &  &  & $-$0.004 &  \\ 
  &  &  & (0.006) &  \\ 

 Gender = Non-binary/Prefer not to respond  &  &  & $-$0.127 &  \\ 
  &  &  & (0.422) &  \\ 

 Gender = Woman &  &  & $-$0.001 &  \\ 
  &  &  & (0.171) &  \\ 

 Race = Black/African American &  &  & $-$0.500 &  \\ 
  &  &  & (0.312) &  \\ 

 Race = Other/Mixed/Prefer not to respond &  &  & $-$0.362 &  \\ 
  &  &  & (0.330) &  \\ 

 Race = White &  &  & $-$0.120 &  \\ 
  &  &  & (0.263) &  \\ 

 Number of images images generated before final selection &  &  &  & $-$0.003 \\ 
  &  &  &  & (0.007) \\ 

 Time until selecting image &  &  &  & $-$0.00002 \\ 
  &  &  &  & (0.0001) \\ 

 Length of prompt (in characters) &  &  &  & $-$0.0005 \\ 
  &  &  &  & (0.001) \\ 

 Image type = Landscape &  &  &  & $-$0.029 \\ 
  &  &  &  & (0.096) \\ 

 image type = Portrait &  &  &  & $-$0.163 \\ 
  &  &  &  & (0.108) \\ 

 Constant & 0.389$^{*}$ & 0.389$^{*}$ & 0.290 & 0.519$^{**}$ \\ 
  & (0.179) & (0.179) & (0.406) & (0.197) \\

\hline
\hline 
\end{tabular} 
  \caption{\textbf{Robustness tests of our results on AI model authorship.} We include three sets of covariates and find consistent experimental effects. Model (1) is the baseline model without any covariates; model (2) includes a variable indicating the order in which the image was shown to an evaluator to identity potential order effects; model (3) includes participant-level covariates; and model (4) includes image-level covariates. Refer to Section~\ref{sec:analysis_plan} for details. Standard errors are shown inside parentheses. $^{*}$p$<$0.05; $^{**}$p$<$0.01; $^{***}$p$<$0.001.} 
  \label{tab:robustness_AI} 
\end{table*} 

\begin{table*}[!htbp] \centering 
\small
\begin{tabular}{@{\extracolsep{5pt}}lcccc} 
\\[-1.8ex]\hline 
\hline \\[-1.8ex] 
 & \multicolumn{4}{c}{\textit{Company Authorship}} \\ 
\cline{2-5} 
\\[-1.8ex] & (1) & (2) & (3) & (4)\\ 
\hline \\[-1.8ex] 
  Invested Evaluators (Own) & $-$0.134 & $-$0.134 & $-$0.209 & $-$0.104 \\ 
  & (0.221) & (0.221) & (0.228) & (0.223) \\ 

 Invested Evaluators (Others) & 0.003 & 0.039 & $-$0.070 & 0.002 \\ 
  & (0.217) & (0.233) & (0.223) & (0.218) \\ 

 Uninvested Evaluators (Others)  & $-$0.132 & $-$0.111 & $-$0.217 & $-$0.128 \\ 
  & (0.213) & (0.226) & (0.209) & (0.213) \\ 

 Evaluation order &  & $-$0.014 &  &  \\ 
  &  & (0.024) &  &  \\ 

 Frequency of using GenAI for images &  &  & $-$0.004 &  \\ 
  &  &  & (0.104) &  \\ 

 Frequency of using GenAI for text &  &  & 0.304$^{**}$ &  \\ 
  &  &  & (0.104) &  \\ 

 Previous participation in GenAI generation study &  &  & 0.367 &  \\ 
  &  &  & (0.251) &  \\ 

 Previous participation in GenAI evaluation study &  &  & 0.118 &  \\ 
  &  &  & (0.237) &  \\ 

 Art training &  &  & $-$0.197 &  \\ 
  &  &  & (0.198) &  \\ 

 Computer science training &  &  & 0.129 &  \\ 
  &  &  & (0.221) &  \\ 

 Law training &  &  & 0.193 &  \\ 
  &  &  & (0.309) &  \\ 

 Age &  &  & 0.004 &  \\ 
  &  &  & (0.006) &  \\ 

 Gender = Non-binary/Prefer not to respond  &  &  & $-$0.136 &  \\ 
  &  &  & (0.474) &  \\ 

 Gender = Woman &  &  & $-$0.025 &  \\ 
  &  &  & (0.169) &  \\ 

 Race = Black/African American &  &  & $-$0.222 &  \\ 
  &  &  & (0.327) &  \\ 

 Race = Other/Mixed/Prefer not to respond &  &  & $-$0.351 &  \\ 
  &  &  & (0.349) &  \\ 

 Race = White &  &  & 0.001 &  \\ 
  &  &  & (0.282) &  \\ 

 Number of images images generated before final selection &  &  &  & 0.013 \\ 
  &  &  &  & (0.011) \\ 

 Time until selecting image &  &  &  & $-$0.0002 \\ 
  &  &  &  & (0.0002) \\ 

 Length of prompt (in characters) &  &  &  & $-$0.0004 \\ 
  &  &  &  & (0.001) \\ 

 Image type = Landscape &  &  &  & 0.050 \\ 
  &  &  &  & (0.105) \\ 

 image type = Portrait &  &  &  & $-$0.058 \\ 
  &  &  &  & (0.107) \\ 

 Constant & $-$0.411$^{*}$ & $-$0.411$^{*}$ & $-$0.956$^{*}$ & $-$0.352 \\ 
  & (0.181) & (0.181) & (0.444) & (0.215) \\

\hline
\hline 
\end{tabular} 
  \caption{\textbf{Robustness tests of our results on company authorship.} We include three sets of covariates and find consistent experimental effects. Model (1) is the baseline model without any covariates; model (2) includes a variable indicating the order in which the image was shown to an evaluator to identity potential order effects; model (3) includes participant-level covariates; and model (4) includes image-level covariates. Refer to Section~\ref{sec:analysis_plan} for details. Standard errors are shown inside parentheses. $^{*}$p$<$0.05; $^{**}$p$<$0.01; $^{***}$p$<$0.001.} 
  \label{tab:robustness_company} 
\end{table*}

\begin{table*}[!htbp] \centering 
\small
\begin{tabular}{@{\extracolsep{5pt}}lcccc} 
\\[-1.8ex]\hline 
\hline \\[-1.8ex] 
 & \multicolumn{4}{c}{\textit{Data Contributors' Authorship}} \\ 
\cline{2-5} 
\\[-1.8ex] & (1) & (2) & (3) & (4)\\ 
\hline \\[-1.8ex] 
 Invested Evaluators (Own) & $-$0.075 & $-$0.075 & 0.089 & $-$0.070 \\ 
  & (0.210) & (0.210) & (0.212) & (0.212) \\ 

 Invested Evaluators (Others) & $-$0.049 & $-$0.054 & 0.107 & $-$0.048 \\ 
  & (0.204) & (0.206) & (0.205) & (0.204) \\ 

 Uninvested Evaluators (Others)  & 0.044 & 0.041 & 0.100 & 0.043 \\ 
  & (0.210) & (0.207) & (0.211) & (0.211) \\ 

 Evaluation order &  & 0.002 &  &  \\ 
  &  & (0.029) &  &  \\ 

 Frequency of using GenAI for images &  &  & $-$0.251$^{*}$ &  \\ 
  &  &  & (0.102) &  \\ 

 Frequency of using GenAI for text &  &  & 0.087 &  \\ 
  &  &  & (0.095) &  \\ 

 Previous participation in GenAI generation study &  &  & $-$0.134 &  \\ 
  &  &  & (0.221) &  \\ 

 Previous participation in GenAI evaluation study &  &  & 0.334 &  \\ 
  &  &  & (0.231) &  \\ 

 Art training &  &  & $-$0.001 &  \\ 
  &  &  & (0.186) &  \\ 

 Computer science training &  &  & 0.425$^{*}$ &  \\ 
  &  &  & (0.216) &  \\ 

 Law training &  &  & 0.177 &  \\ 
  &  &  & (0.274) &  \\ 

 Age &  &  & $-$0.017$^{**}$ &  \\ 
  &  &  & (0.006) &  \\ 

 Gender = Non-binary/Prefer not to respond  &  &  & $-$0.774 &  \\ 
  &  &  & (0.500) &  \\ 

 Gender = Woman &  &  & 0.511$^{**}$ &  \\ 
  &  &  & (0.161) &  \\ 

 Race = Black/African American &  &  & 0.092 &  \\ 
  &  &  & (0.322) &  \\ 

 Race = Other/Mixed/Prefer not to respond &  &  & 0.562 &  \\ 
  &  &  & (0.333) &  \\ 

 Race = White &  &  & 0.205 &  \\ 
  &  &  & (0.268) &  \\ 

 Number of images images generated before final selection &  &  &  & 0.0005 \\ 
  &  &  &  & (0.012) \\ 

 Time until selecting image &  &  &  & $-$0.0001 \\ 
  &  &  &  & (0.0002) \\ 

 Length of prompt (in characters) &  &  &  & 0.001 \\ 
  &  &  &  & (0.0005) \\ 

 Image type = Landscape &  &  &  & 0.081 \\ 
  &  &  &  & (0.098) \\ 

 image type = Portrait &  &  &  & 0.106 \\ 
  &  &  &  & (0.113) \\ 

 Constant & 0.589$^{***}$ & 0.589$^{***}$ & 0.741 & 0.493$^{**}$ \\ 
  & (0.167) & (0.167) & (0.412) & (0.176) \\

\hline
\hline 
\end{tabular} 
  \caption{\textbf{Robustness tests of our results on data contributors' authorship.} We include three sets of covariates and find consistent experimental effects. Model (1) is the baseline model without any covariates; model (2) includes a variable indicating the order in which the image was shown to an evaluator to identity potential order effects; model (3) includes participant-level covariates; and model (4) includes image-level covariates. Refer to Section~\ref{sec:analysis_plan} for details. Standard errors are shown inside parentheses. $^{*}$p$<$0.05; $^{**}$p$<$0.01; $^{***}$p$<$0.001.} 
  \label{tab:robustness_artists} 
\end{table*} 

\begin{table*}[!htbp] \centering 
\small
\begin{tabular}{@{\extracolsep{5pt}}lcccc} 
\\[-1.8ex]\hline 
\hline \\[-1.8ex] 
 & \multicolumn{4}{c}{\textit{Score Evaluation}} \\ 
\cline{2-5} 
\\[-1.8ex] & (1) & (2) & (3) & (4)\\ 
\hline \\[-1.8ex] 
 Invested Evaluators (Own) & $-$0.178 & $-$0.178 & $-$0.333 & $-$0.122 \\ 
  & (0.217) & (0.217) & (0.226) & (0.226) \\ 

 Invested Evaluators (Others) & $-$1.743$^{***}$ & $-$1.510$^{***}$ & $-$1.894$^{***}$ & $-$1.723$^{***}$ \\ 
  & (0.213) & (0.222) & (0.225) & (0.213) \\ 

 Uninvested Evaluators (Others)  & $-$1.986$^{***}$ & $-$1.847$^{***}$ & $-$2.035$^{***}$ & $-$2.004$^{***}$ \\ 
  & (0.225) & (0.226) & (0.234) & (0.225) \\ 

 Evaluation order &  & $-$0.094$^{*}$ &  &  \\ 
  &  & (0.046) &  &  \\ 

 Frequency of using GenAI for images &  &  & 0.397$^{***}$ &  \\ 
  &  &  & (0.098) &  \\ 

 Frequency of using GenAI for text &  &  & 0.111 &  \\ 
  &  &  & (0.089) &  \\ 

 Previous participation in GenAI generation study &  &  & 0.137 &  \\ 
  &  &  & (0.228) &  \\ 

 Previous participation in GenAI evaluation study &  &  & $-$0.299 &  \\ 
  &  &  & (0.255) &  \\ 

 Art training &  &  & $-$0.114 &  \\ 
  &  &  & (0.195) &  \\ 

 Computer science training &  &  & $-$0.045 &  \\ 
  &  &  & (0.198) &  \\ 

 Law training &  &  & $-$0.616$^{*}$ &  \\ 
  &  &  & (0.312) &  \\ 

 Age &  &  & 0.009 &  \\ 
  &  &  & (0.006) &  \\ 

 Gender = Non-binary/Prefer not to respond  &  &  & 0.075 &  \\ 
  &  &  & (0.617) &  \\ 

 Gender = Woman &  &  & $-$0.126 &  \\ 
  &  &  & (0.165) &  \\ 

 Race = Black/African American &  &  & 0.063 &  \\ 
  &  &  & (0.332) &  \\ 

 Race = Other/Mixed/Prefer not to respond &  &  & 0.166 &  \\ 
  &  &  & (0.320) &  \\ 

 Race = White &  &  & 0.182 &  \\ 
  &  &  & (0.251) &  \\ 

 Number of images images generated before final selection &  &  &  & 0.028 \\ 
  &  &  &  & (0.017) \\ 

 Time until selecting image &  &  &  & 0.0001 \\ 
  &  &  &  & (0.0002) \\ 

 Length of prompt (in characters) &  &  &  & 0.003$^{***}$ \\ 
  &  &  &  & (0.001) \\ 

 Image type = Landscape &  &  &  & 0.440$^{*}$ \\ 
  &  &  &  & (0.194) \\ 

 image type = Portrait &  &  &  & 0.405 \\ 
  &  &  &  & (0.220) \\ 

 Constant & 8.189$^{***}$ & 8.189$^{***}$ & 7.350$^{***}$ & 7.395$^{***}$ \\ 
  & (0.163) & (0.163) & (0.387) & (0.240) \\

\hline
\hline 
\end{tabular} 
  \caption{\textbf{Robustness tests of our score evaluation results.} We include three sets of covariates and find consistent experimental effects. Model (1) is the baseline model without any covariates; model (2) includes a variable indicating the order in which the image was shown to an evaluator to identity potential order effects; model (3) includes participant-level covariates; and model (4) includes image-level covariates. Refer to Section~\ref{sec:analysis_plan} for details. Standard errors are shown inside parentheses. $^{*}$p$<$0.05; $^{**}$p$<$0.01; $^{***}$p$<$0.001.} 
  \label{tab:robustness_exhibition} 
\end{table*} 

\newpage



\begin{table*}[!htbp] \centering 
\small
\begin{tabular}{@{\extracolsep{5pt}} l|rrlr} 
\\[-1.8ex]\hline 
\hline \\[-1.8ex]
Contrast & OR & SE & z-test & p-value \\ 
\hline \\[-1.8ex] 
\textbf{User} & & & & \\
Creators (Own) / Uninvested Evaluators (Others) / H1 and H4 & $0.774$ & $0.189$ & z = -1.048 & $1.000$ \\ 
Creators (Own) / Invested Evaluators (Others) / H1 & $0.635$ & $0.161$ & z = -1.786 & $0.371$ \\ 
Invested Evaluators (Own) / Invested Evaluators (Others) / H2 & $0.884$ & $0.082$ & z = -1.330 & $0.917$ \\ 
Uninvested Evaluators (Others) / Invested Evaluators (Others) / H3 and H4 & $0.820$ & $0.169$ & z = -0.959 & $1.000$ \\ 
Invested Evaluators (Own) / Uninvested Evaluators (Others) /  H4 & $1.077$ & $0.230$ & z = 0.348 & $1.000$ \\ 
\\[-1.8ex] 
\textbf{AI Model} & & & & \\
Creators (Own) / Uninvested Evaluators (Others) / H1 and H4 & $1.400$ & $0.413$ & z = 1.140 & $1.000$ \\ 
Creators (Own) / Invested Evaluators (Others) / H1 & $1.078$ & $0.318$ & z = 0.254 & $1.000$ \\ 
Invested Evaluators (Own) / Invested Evaluators (Others) / H2 & $1.259$ & $0.141$ & z = 2.052 & $0.201$ \\ 
Uninvested Evaluators (Others) / Invested Evaluators (Others) / H3 and H4 & $0.770$ & $0.187$ & z = -1.075 & $1.000$ \\ 
Invested Evaluators (Own) / Uninvested Evaluators (Others) /  H4 & $1.635$ & $0.404$ & z = 1.992 & $0.232$ \\ 
\\[-1.8ex] 
\textbf{Company} & & & & \\
Creators (Own) / Uninvested Evaluators (Others) / H1 and H4 & $1.333$ & $0.361$ & z = 1.062 & $1.000$ \\ 
Creators (Own) / Invested Evaluators (Others) / H1 & $0.816$ & $0.221$ & z = -0.751 & $1.000$ \\ 
Invested Evaluators (Own) / Invested Evaluators (Others) / H2 & $1.055$ & $0.086$ & z = 0.659 & $1.000$ \\ 
Uninvested Evaluators (Others) / Invested Evaluators (Others) / H3 and H4 & $0.612$ & $0.130$ & z = -2.308 & $0.105$ \\ 
Invested Evaluators (Own) / Uninvested Evaluators (Others) /  H4 & $1.724$ & $0.381$ & z = 2.464 & $0.069$ \\
\\[-1.8ex] 
\textbf{Data Contributors} & & & & \\
Creators (Own) / Uninvested Evaluators (Others) / H1 and H4 & $0.566$ & $0.147$ & z = -2.189 & $0.143$ \\ 
Creators (Own) / Invested Evaluators (Others) / H1 & $0.710$ & $0.174$ & z = -1.396 & $0.814$ \\ 
Invested Evaluators (Own) / Invested Evaluators (Others) / H2 & $1.007$ & $0.095$ & z = 0.075 & $1.000$ \\ 
Uninvested Evaluators (Others) / Invested Evaluators (Others) / H3 and H4 & $1.254$ & $0.255$ & z = 1.112 & $1.000$ \\ 
Invested Evaluators (Own) / Uninvested Evaluators (Others) /  H4 & $0.803$ & $0.167$ & z = -1.056 & $1.000$ \\
\\[-1.8ex] 
\textbf{Anyone} & & & & \\
Creators (Own) / Uninvested Evaluators (Others) / H1 and H4 & $1.520$ & $0.482$ & z = 1.321 & $0.933$ \\ 
Creators (Own) / Invested Evaluators (Others) / H1 & $3.953$ & $1.631$ & z = 3.330 & \bm{$0.004$} \\ 
Invested Evaluators (Own) / Invested Evaluators (Others) / H2 & $1.763$ & $0.471$ & z = 2.122 & $0.169$ \\ 
Uninvested Evaluators (Others) / Invested Evaluators (Others) / H3 and H4 & $2.600$ & $0.981$ & z = 2.532 & $0.057$ \\ 
Invested Evaluators (Own) / Uninvested Evaluators (Others) /  H4 & $0.678$ & $0.236$ & z = -1.114 & $1.000$ \\
\hline
\end{tabular} 
  \caption{\textbf{Right to display commercially:} Pairwise comparisons of attributions of the right to display AI-generated images commercially between treatment conditions. We only test the contrasts relevant to our hypotheses presented in Section~\ref{sec:hypotheses}. We apply Bonferroni corrections to account for multiple comparisons. OR refers to the odds ratio.} 
  \label{tab:dp_pairwise} 
\end{table*}


\begin{table*}[!htbp] \centering 
\small
\begin{tabular}{@{\extracolsep{5pt}} l|rrlr} 
\\[-1.8ex]\hline 
\hline \\[-1.8ex]
Contrast & OR & SE & z-test & p-value \\ 
\hline \\[-1.8ex] 
\textbf{User} & & & & \\
Creators (Own) / Uninvested Evaluators (Others) / H1 and H4 & $1.251$ & $0.323$ & z = 0.867 & $1.000$ \\ 
Creators (Own) / Invested Evaluators (Others) / H1 & $0.812$ & $0.212$ & z = -0.795 & $1.000$ \\ 
Invested Evaluators (Own) / Invested Evaluators (Others) / H2 & $1.059$ & $0.102$ & z = 0.596 & $1.000$ \\ 
Uninvested Evaluators (Others) / Invested Evaluators (Others) / H3 and H4 & $0.649$ & $0.139$ & z = -2.011 & $0.222$ \\ 
Invested Evaluators (Own) / Uninvested Evaluators (Others) /  H4 & $1.631$ & $0.370$ & z = 2.156 & $0.155$ \\ 
\\[-1.8ex] 
\textbf{AI Model} & & & & \\
Creators (Own) / Uninvested Evaluators (Others) / H1 and H4 & $1.447$ & $0.411$ & z = 1.302 & $0.964$ \\ 
Creators (Own) / Invested Evaluators (Others) / H1 & $1.146$ & $0.306$ & z = 0.508 & $1.000$ \\ 
Invested Evaluators (Own) / Invested Evaluators (Others) / H2 & $1.204$ & $0.126$ & z = 1.774 & $0.380$ \\ 
Uninvested Evaluators (Others) / Invested Evaluators (Others) / H3 and H4 & $0.792$ & $0.177$ & z = -1.044 & $1.000$ \\ 
Invested Evaluators (Own) / Uninvested Evaluators (Others) /  H4 & $1.520$ & $0.356$ & z = 1.792 & $0.366$ \\
\\[-1.8ex] 
\textbf{Company} & & & & \\
Creators (Own) / Uninvested Evaluators (Others) / H1 and H4 & $1.300$ & $0.340$ & z = 1.005 & $1.000$ \\ 
Creators (Own) / Invested Evaluators (Others) / H1 & $1.051$ & $0.266$ & z = 0.198 & $1.000$ \\ 
Invested Evaluators (Own) / Invested Evaluators (Others) / H2 & $1.064$ & $0.095$ & z = 0.695 & $1.000$ \\ 
Uninvested Evaluators (Others) / Invested Evaluators (Others) / H3 and H4 & $0.808$ & $0.166$ & z = -1.037 & $1.000$ \\ 
Invested Evaluators (Own) / Uninvested Evaluators (Others) /  H4 & $1.316$ & $0.283$ & z = 1.277 & $1.000$ \\
\\[-1.8ex] 
\textbf{Data Contributors} & & & & \\
Creators (Own) / Uninvested Evaluators (Others) / H1 and H4 & $0.708$ & $0.183$ & z = -1.334 & $0.911$ \\ 
Creators (Own) / Invested Evaluators (Others) / H1 & $0.797$ & $0.208$ & z = -0.867 & $1.000$ \\ 
Invested Evaluators (Own) / Invested Evaluators (Others) / H2 & $1.068$ & $0.102$ & z = 0.684 & $1.000$ \\ 
Uninvested Evaluators (Others) / Invested Evaluators (Others) / H3 and H4 & $1.127$ & $0.233$ & z = 0.577 & $1.000$ \\ 
Invested Evaluators (Own) / Uninvested Evaluators (Others) /  H4 & $0.948$ & $0.200$ & z = -0.254 & $1.000$ \\ 
\\[-1.8ex] 
\textbf{Anyone} & & & & \\
Creators (Own) / Uninvested Evaluators (Others) / H1 and H4 & $1.666$ & $0.463$ & z = 1.835 & $0.333$ \\ 
Creators (Own) / Invested Evaluators (Others) / H1 & $2.193$ & $0.585$ & z = 2.941 & \bm{$0.016$} \\ 
Invested Evaluators (Own) / Invested Evaluators (Others) / H2 & $1.118$ & $0.096$ & z = 1.297 & $0.973$ \\ 
Uninvested Evaluators (Others) / Invested Evaluators (Others) / H3 and H4 & $1.316$ & $0.277$ & z = 1.304 & $0.961$ \\ 
Invested Evaluators (Own) / Uninvested Evaluators (Others) /  H4 & $0.849$ & $0.187$ & z = -0.742 & $1.000$ \\ 
\hline
\end{tabular} 
  \caption{\textbf{Right to display non-commercially:} Pairwise comparisons of attributions of the right to display AI-generated images non-commercially between treatment conditions. We only test the contrasts relevant to our hypotheses presented in Section~\ref{sec:hypotheses}. We apply Bonferroni corrections to account for multiple comparisons. OR refers to the odds ratio.} 
  \label{tab:dn_pairwise} 
\end{table*}


\begin{table*}[!htbp] \centering 
\small
\begin{tabular}{@{\extracolsep{5pt}} l|rrlr} 
\\[-1.8ex]\hline 
\hline \\[-1.8ex]
Contrast & OR & SE & z-test & p-value \\ 
\hline \\[-1.8ex] 
\textbf{User} & & & & \\
Creators (Own) / Uninvested Evaluators (Others) / H1 and H4 & $0.898$ & $0.226$ & z = -0.429 & $1.000$ \\ 
Creators (Own) / Invested Evaluators (Others) / H1 & $0.774$ & $0.197$ & z = -1.008 & $1.000$ \\ 
Invested Evaluators (Own) / Invested Evaluators (Others) / H2 & $1.049$ & $0.096$ & z = 0.525 & $1.000$ \\ 
Uninvested Evaluators (Others) / Invested Evaluators (Others) / H3 and H4 & $0.862$ & $0.179$ & z = -0.714 & $1.000$ \\ 
Invested Evaluators (Own) / Uninvested Evaluators (Others) /  H4 & $1.217$ & $0.264$ & z = 0.906 & $1.000$ \\
\\[-1.8ex] 
\textbf{AI Model} & & & & \\
Creators (Own) / Uninvested Evaluators (Others) / H1 and H4 & $1.033$ & $0.351$ & z = 0.096 & $1.000$ \\ 
Creators (Own) / Invested Evaluators (Others) / H1 & $0.832$ & $0.279$ & z = -0.549 & $1.000$ \\ 
Invested Evaluators (Own) / Invested Evaluators (Others) / H2 & $1.117$ & $0.116$ & z = 1.069 & $1.000$ \\ 
Uninvested Evaluators (Others) / Invested Evaluators (Others) / H3 and H4 & $0.805$ & $0.215$ & z = -0.812 & $1.000$ \\ 
Invested Evaluators (Own) / Uninvested Evaluators (Others) /  H4 & $1.388$ & $0.377$ & z = 1.206 & $1.000$ \\  
\\[-1.8ex] 
\textbf{Company} & & & & \\
Creators (Own) / Uninvested Evaluators (Others) / H1 and H4 & $0.997$ & $0.277$ & z = -0.011 & $1.000$ \\ 
Creators (Own) / Invested Evaluators (Others) / H1 & $0.674$ & $0.187$ & z = -1.424 & $0.772$ \\ 
Invested Evaluators (Own) / Invested Evaluators (Others) / H2 & $0.985$ & $0.082$ & z = -0.181 & $1.000$ \\ 
Uninvested Evaluators (Others) / Invested Evaluators (Others) / H3 and H4 & $0.676$ & $0.146$ & z = -1.811 & $0.351$ \\ 
Invested Evaluators (Own) / Uninvested Evaluators (Others) /  H4 & $1.456$ & $0.328$ & z = 1.671 & $0.474$ \\ 
\\[-1.8ex] 
\textbf{Data Contributors} & & & & \\
Creators (Own) / Uninvested Evaluators (Others) / H1 and H4 & $0.502$ & $0.134$ & z = -2.589 & \bm{$0.048$} \\ 
Creators (Own) / Invested Evaluators (Others) / H1 & $0.742$ & $0.191$ & z = -1.158 & $1.000$ \\ 
Invested Evaluators (Own) / Invested Evaluators (Others) / H2 & $1.046$ & $0.085$ & z = 0.551 & $1.000$ \\ 
Uninvested Evaluators (Others) / Invested Evaluators (Others) / H3 and H4 & $1.478$ & $0.304$ & z = 1.898 & $0.289$ \\ 
Invested Evaluators (Own) / Uninvested Evaluators (Others) /  H4 & $0.708$ & $0.148$ & z = -1.659 & $0.486$ \\  
\\[-1.8ex] 
\textbf{Anyone} & & & & \\
Creators (Own) / Uninvested Evaluators (Others) / H1 and H4 & $1.449$ & $0.511$ & z = 1.052 & $1.000$ \\ 
Creators (Own) / Invested Evaluators (Others) / H1 & $1.863$ & $0.698$ & z = 1.660 & $0.485$ \\ 
Invested Evaluators (Own) / Invested Evaluators (Others) / H2 & $1.050$ & $0.189$ & z = 0.270 & $1.000$ \\ 
Uninvested Evaluators (Others) / Invested Evaluators (Others) / H3 and H4 & $1.286$ & $0.420$ & z = 0.770 & $1.000$ \\ 
Invested Evaluators (Own) / Uninvested Evaluators (Others) /  H4 & $0.816$ & $0.286$ & z = -0.579 & $1.000$ \\
\hline
\end{tabular} 
  \caption{\textbf{Right to make copies commercially:} Pairwise comparisons of attributions of the right to make copies of AI-generated images commercially between treatment conditions. We only test the contrasts relevant to our hypotheses presented in Section~\ref{sec:hypotheses}. We apply Bonferroni corrections to account for multiple comparisons. OR refers to the odds ratio.} 
  \label{tab:cp_pairwise} 
\end{table*}


\begin{table*}[!htbp] \centering 
\small
\begin{tabular}{@{\extracolsep{5pt}} l|rrlr} 
\\[-1.8ex]\hline 
\hline \\[-1.8ex]
Contrast & OR & SE & z-test & p-value \\ 
\hline \\[-1.8ex] 
\textbf{User} & & & & \\
Creators (Own) / Uninvested Evaluators (Others) / H1 and H4 & $1.372$ & $0.362$ & z = 1.199 & $1.000$ \\ 
Creators (Own) / Invested Evaluators (Others) / H1 & $0.905$ & $0.243$ & z = -0.37 & $1.000$ \\ 
Invested Evaluators (Own) / Invested Evaluators (Others) / H2 & $1.033$ & $0.106$ & z = 0.314 & $1.000$ \\ 
Uninvested Evaluators (Others) / Invested Evaluators (Others) / H3 and H4 & $0.660$ & $0.139$ & z = -1.968 & $0.245$ \\ 
Invested Evaluators (Own) / Uninvested Evaluators (Others) /  H4 & $1.565$ & $0.354$ & z = 1.979 & $0.239$ \\
\\[-1.8ex] 
\textbf{AI Model} & & & & \\
Creators (Own) / Uninvested Evaluators (Others) / H1 and H4 & $1.667$ & $0.468$ & z = 1.818 & $0.345$ \\ 
Creators (Own) / Invested Evaluators (Others) / H1 & $1.307$ & $0.344$ & z = 1.016 & $1.000$ \\ 
Invested Evaluators (Own) / Invested Evaluators (Others) / H2 & $0.934$ & $0.118$ & z = -0.544 & $1.000$ \\ 
Uninvested Evaluators (Others) / Invested Evaluators (Others) / H3 and H4 & $0.784$ & $0.176$ & z = -1.084 & $1.000$ \\ 
Invested Evaluators (Own) / Uninvested Evaluators (Others) /  H4 & $1.191$ & $0.286$ & z = 0.729 & $1.000$ \\ 
\\[-1.8ex] 
\textbf{Company} & & & & \\
Creators (Own) / Uninvested Evaluators (Others) / H1 and H4 & $1.365$ & $0.355$ & z = 1.197 & $1.000$ \\ 
Creators (Own) / Invested Evaluators (Others) / H1 & $0.978$ & $0.247$ & z = -0.089 & $1.000$ \\ 
Invested Evaluators (Own) / Invested Evaluators (Others) / H2 & $0.956$ & $0.084$ & z = -0.511 & $1.000$ \\ 
Uninvested Evaluators (Others) / Invested Evaluators (Others) / H3 and H4 & $0.716$ & $0.146$ & z = -1.641 & $0.504$ \\ 
Invested Evaluators (Own) / Uninvested Evaluators (Others) /  H4 & $1.335$ & $0.286$ & z = 1.347 & $0.890$ \\
\\[-1.8ex] 
\textbf{Data Contributors} & & & & \\
Creators (Own) / Uninvested Evaluators (Others) / H1 and H4 & $0.712$ & $0.187$ & z = -1.289 & $0.987$ \\ 
Creators (Own) / Invested Evaluators (Others) / H1 & $0.855$ & $0.220$ & z = -0.611 & $1.000$ \\ 
Invested Evaluators (Own) / Invested Evaluators (Others) / H2 & $0.903$ & $0.095$ & z = -0.977 & $1.000$ \\ 
Uninvested Evaluators (Others) / Invested Evaluators (Others) / H3 and H4 & $1.200$ & $0.248$ & z = 0.882 & $1.000$ \\ 
Invested Evaluators (Own) / Uninvested Evaluators (Others) /  H4 & $0.752$ & $0.160$ & z = -1.340 & $0.901$ \\
\\[-1.8ex] 
\textbf{Anyone} & & & & \\
Creators (Own) / Uninvested Evaluators (Others) / H1 and H4 & $1.037$ & $0.265$ & z = 0.142 & $1.000$ \\ 
Creators (Own) / Invested Evaluators (Others) / H1 & $1.477$ & $0.367$ & z = 1.571 & $0.581$ \\ 
Invested Evaluators (Own) / Invested Evaluators (Others) / H2 & $0.987$ & $0.095$ & z = -0.137 & $1.000$ \\ 
Uninvested Evaluators (Others) / Invested Evaluators (Others) / H3 and H4 & $1.425$ & $0.306$ & z = 1.648 & $0.497$ \\ 
Invested Evaluators (Own) / Uninvested Evaluators (Others) /  H4 & $0.693$ & $0.149$ & z = -1.703 & $0.443$ \\
\hline
\end{tabular} 
  \caption{\textbf{Right to make copies non-commercially:} Pairwise comparisons of attributions of the right to make copies of AI-generated images non-commercially between treatment conditions. We only test the contrasts relevant to our hypotheses presented in Section~\ref{sec:hypotheses}. We apply Bonferroni corrections to account for multiple comparisons. OR refers to the odds ratio.} 
  \label{tab:cn_pairwise} 
\end{table*}

\end{document}